\pdfoutput=1
\documentclass[11pt]{article}
\usepackage{amssymb}
\usepackage{amsmath}
\usepackage{epsfig}
\usepackage{graphicx}

\usepackage[inner=2.5cm,outer=2cm]{geometry} 
\usepackage{bm}
\usepackage{caption}
\usepackage{subcaption}
\usepackage{sidecap}
\newcommand{\N}{\mathcal{N}}

\newcommand{\R}{\mathbb{R}}

\newcommand{\argmax}{\operatornamewithlimits{argmax}}

\newcommand{\E}{\text{E}}

\long\def\comment#1{}

\title{Robust and scalable Bayesian analysis of spatial neural tuning function data}
\author{Kamiar Rahnama Rad\thanks{KR was supported by the PSC-CUNY Award  67377-00 45.} \and Timothy A. Machado \thanks{TAM was supported by the NSF GRFP.} \and  Liam Paninski \thanks{LP was supported by the Gatsby Foundation, NSF CAREER award, and grants ARO MURI W911NF-12-1-0594, NSF CAREER IOS-0641912, ONR N00014-14-1-0243, and DARPA N66001-15-C-4032.}}
\date{{City University of New York and Columbia University}}

\begin{document}

\maketitle

\begin{abstract}
A common analytical problem in neuroscience is the interpretation of neural activity with respect to sensory input or behavioral output. This is typically achieved by regressing measured neural activity against known stimuli or behavioral variables to produce a ``tuning function" for each neuron. Unfortunately, because this approach handles neurons individually, it cannot take advantage of simultaneous measurements from spatially adjacent neurons that often have similar tuning properties. On the other hand, sharing information between adjacent neurons can errantly degrade estimates of tuning functions across space if there are sharp discontinuities in tuning between nearby neurons. In this paper, we develop a computationally efficient block Gibbs sampler that effectively pools information between neurons to de-noise tuning function estimates while simultaneously preserving sharp discontinuities that might exist in the organization of tuning across space. This method is fully Bayesian and its computational cost per iteration scales sub-quadratically with total parameter dimensionality. We demonstrate the robustness and scalability of this approach by applying it to both real and synthetic datasets.  In particular, an application to data from the spinal cord illustrates that the proposed methods can dramatically decrease the experimental time required to accurately estimate tuning functions.
\end{abstract}

\section{Introduction}
{Over the past five years, it has become possible to simultaneously record the activity of thousands of neurons at single-cell resolution \cite{AORLK13,PYH14,PFEO14,HGPS15}.} The high spatial and temporal resolution permitted by these new methods allows us to examine whether previously unexamined regions of the brain might dynamically map sensory information across space in unappreciated ways. However, the high dimensionality of these data also poses new computational challenges for statistical neuroscientists. Therefore scalable and efficient methods for extracting as much information as possible from these recordings must be developed; {in turn, improved analytical approaches that can extract information from e.g. shorter experiments may enable new dynamic closed-loop experimental designs}. 

In many experimental settings, a key quantity of interest is the tuning function, a filter that relates known information about sensory input or behavioral state to the activity of a neuron. For example, tuning functions permit measurement of orientation
selectivity in visual cortex \cite{HW68}, allow us to relate movement direction to activity in primary motor cortex
\cite{S00,GCK86}, and let us measure the grid-like spatial sensitivity of neurons within entorhinal cortex \cite{HFMMM05}.  This paper focuses on data-efficient methods for tuning function estimation.

To be more concrete, let us first consider example experimental data where the activity of $n$ neurons is measured across $d$ trials of identical lengths, with different stimuli presented during each trial. We can then model the response $\bm{y_i} \in \R^d$ of neuron $i$ as a function of a stimulus matrix $\bm{X_i} \in \R^{d \times m}$. Each row of $\bm{X_i}$ corresponds to the stimulus projected onto neuron $i$, at each of the $d$ trials. In the simplest case, the relationship between the unobserved tuning function $\bm{\beta_i} \in \R^m$ and the observed activity $\bm{y_i}$ at neuron $i$ in response to stimulus $\bm{X_i}$ can be modeled as\footnote{ {Empirical findings, to some degree, challenge the linear neural response to the stimulus,  the conditionally independent neural activity,  and the Gaussian noise  assumptions.  Nevertheless, numerous studies have successfully used these simplifying assumptions to analyze neural data (see \cite{RWRB97, DS12} and references therein). In the concluding section \ref{sec:cr}, we discuss directions for future work that allow  the approach presented here to be extended to more general settings, e.g. correlated point process observations.}}: 
\begin{eqnarray}
\bm{y_i =  X_i  \beta_i + \epsilon_i} \text{         where    } \bm{\epsilon_i} \sim \N(0,\nu_i^2\sigma^2 \bm{I}). \label{eq:model}
\end{eqnarray}
The efficient statistical analysis and estimation of the unobserved tuning functions $\{ \bm{\beta_i} \}$ given the noisy observations $\{ \bm{y_i}\}$ and the stimulus set $\{\bm{X_i}\}$ is the tuning function estimation problem. In this setting, one standard approach is to use, for example, maximum-likelihood estimation to estimate tuning functions one neuron at a time (e.g., $\bm{\beta_{i,\text{ml}}} := \bm{(X_i'X_i)^{-1} X_i' y_i}$). 

However, this model neglects a common feature of many neural circuits: the spatial clustering of neurons sharing a similar information processing function. {For example, there are maps of tone frequency across the cortical surface in the auditory system \cite{issa2014multiscale}, visual orientation maps in both cortical \cite{HW62,HW68,OHKI05} and subcortical brain regions \cite{feinberg2014orientation}, and maps respecting the spatial organization of the body (somatotopy) in the motor system \cite{LS17,penfield1950cerebral,romanes1964motor,bouchard2013functional,MPPJM15}. As a consequence, neurons in close proximity often have similar tuning functions (see \cite{S08,WM15}, for recent reviews).} In each of these cases, there are typically regions where this rule is violated and largely smooth tuning maps are punctuated by jumps or discontinuities. Therefore simply smoothing in all cases will erode the precision of any sharp borders that might exist. Ideally, we would use an approach to estimate $\{ \bm{\beta_i} \}$ that would smooth out the tuning map more in areas where there is evidence from the data that nearby tuning functions are similar, while letting the data `speak for itself' and applying minimal smoothing in regions where adjacent neurons have tuning functions that are very dissimilar. 


In this paper, we propose a multivariate Bayesian extension of group lasso \cite{YL06}, generalized lasso \cite{TT11}, and total-variation (TV) regularization \cite{ROF92}. Specifically, we use the following { improper} prior:
\begin{eqnarray}
{\bm{\beta}| \lambda,\sigma} &\propto& \prod_{i \sim j}\bigl ( \frac{\lambda}{2 \sigma} \bigr)^m \exp \Bigl( -\frac{\lambda}{\sigma} \Bigl \| \bm{ \beta_i - \beta_j }\Bigr\|_2 \Bigr)\label{eq:beta_prior},
\end{eqnarray}
where $\|u\|_2 = \sqrt{\sum_{i=1}^m u_i^2} $ and $i \sim j$ if two cells $i$ and $j$ are spatially nearby\footnote{ We will clearly define the notion of proximity $i \sim j$, at the end of section \ref{sec:ldn}.}. This prior allows for a flexible level of similarity between nearby tuning functions. For clarity, we contrast against a $\| \bm{ \beta_i - \beta_j}\|_2^2$ based prior:
\begin{eqnarray*}
 \prod_{i \sim j}\bigl ( \frac{\lambda^2}{2 \pi \sigma^2} \bigr)^{m/2} \exp \Bigl( -\frac{\lambda^2}{2\sigma^2} \Bigl \| \bm{ \beta_i - \beta_j }\Bigr\|_2^2 \Bigr),
\end{eqnarray*}
which penalizes large local differences quadratically. The prior defined in (\ref{eq:beta_prior}), on the other hand, penalizes large differences linearly; intuitively, this prior encourages nearby tuning functions to be similar while allowing for large occasional breaks or outliers in the spatial map of the inferred  tuning functions.  This makes the estimates much more robust to these occasional breaks.

The paper is organized as follows. Section 2 presents the full description of our statistical model, including likelihood, priors and hyper-priors. Section 3 presents an efficient block Gibbs sampler with discussions about its statistical and computational properties. Finally, section 4 illustrates our robust and scalable Bayesian analysis of simulated data from the visual cortex and real neural data obtained from the spinal cord. We conclude in Section 5 with a discussion of related work and possible extensions to our approach.

\section{Bayesian Inference\label{sec:ldn}}

To complete the model introduced above, we place an inverse Gamma prior on $\sigma$ and $\{ \nu_i\}_{i=1,\cdots,n}$,  and we place a Gamma prior on $\lambda^2$, both of which are fairly common choices in Bayesian inference \cite{PG08}. These choices lead to the likelihood, priors, and hyper-priors presented below:
\begin{eqnarray}
\text{likelihood,}\hspace{0.6cm}  \bm{y_i | \beta_i},\sigma,\nu_i &\sim&  \bigl( \frac{1}{2 \pi \nu_i^2 \sigma^2} \bigr)^{d/2} \exp \Bigl( -\frac{1}{2\nu_i^2 \sigma^2} \Bigl \| \bm{  y_i - X_i \beta_i }\Bigr\|_2^2 \Bigr) 
  \nonumber 
\\
\text{prior,}\hspace{1.35cm}  \bm{\beta} | \lambda,\sigma  &\sim& \prod_{i \sim j}\bigl ( \frac{\lambda}{2 \sigma} \bigr)^m \exp \Bigl( -\frac{\lambda}{\sigma} \Bigl \|  \bm{\beta_i - \beta_j} \Bigr\|_2 \Bigr) \nonumber\end{eqnarray}
and hyper-priors,
\begin{eqnarray}
\sigma^2 &\sim&  \text{inverse-Gamma}(\kappa,\epsilon)=\frac{\epsilon^{\kappa} }{\Gamma(\kappa)} (\sigma^2)^{-\kappa-1} e^{-\epsilon/\sigma^2} \label{eq:hpriors} \\
 \lambda^2 &\sim&  \text{Gamma}(r,\delta)= \frac{\delta^r }{\Gamma(r)} (\lambda^2)^{r-1} e^{-\delta \lambda^2}\nonumber\\
 \nu_i^2 &\sim&   \text{inverse-Gamma}(\varkappa,\varepsilon)=\frac{\varepsilon^{\varkappa} }{\Gamma(\varkappa)} (\nu_i^2)^{-\varkappa-1} e^{-\varepsilon/\nu_i^2}.\nonumber
\end{eqnarray}  

 The well known representation \cite{AM74,W87, ETL06,CGGK10}  of the Laplace prior as a scale mixture of Normals:
 
\begin{eqnarray*}
&&\bigl ( \frac{\lambda}{2 \sigma} \bigr)^m \exp \Bigl( -\frac{\lambda}{\sigma} \|  \bm{\beta_i - \beta_j }\|_2 \Bigr)=\\ && C\int_0^{\infty} \Bigl (\frac{1}{2 \pi \sigma^2 \tau_{ij}^2} \Bigr)^{m/2} \exp \Bigl( -\frac{\| \bm{ \beta_i - \beta_j} \|_2^2}{2\sigma^2 \tau_{ij}^2}  \Bigr) \underbrace{ \frac{(\frac{\lambda^2}{2})^{\frac{m+1}{2}} }{\Gamma(\frac{m+1}{2})}  (\tau_{ij}^2)^{\frac{m+1}{2}-1 } e^{- \frac{\lambda^2}{2} \tau_{ij}^2} d\tau_{ij}^2}_{\tau_{ij}^2 \sim \text{Gamma}(\frac{m+1}{2}, \frac{\lambda^2}{2}) },
\end{eqnarray*}
(where $C=\pi^{\frac{m-1}{2}} \Gamma(\frac{m+1}{2})$) allows us to formulate our prior (\ref{eq:beta_prior}), in a hierarchical manner:
\begin{eqnarray}
\tau_{ij}^2 | \lambda^2 &\sim&  \frac{(\frac{\lambda^2}{2})^{\frac{m+1}{2}} }{\Gamma(\frac{m+1}{2})}  (\tau_{ij}^2)^{\frac{m+1}{2}-1 } e^{- \frac{\lambda^2}{2} \tau_{ij}^2} \quad \text{for all $i\sim j$} \label{eq:t}\\
\bm{\beta}| \{\tau_{ij}^2\},\sigma^2 &\sim& \exp \bigl(-\frac{\bm{\beta' D' \Gamma D \beta}}{2\sigma^2}  \bigr) \label{eq:hie}
\end{eqnarray}
where {(using $\otimes$ as the Kronecker product)}
\begin{eqnarray*}
\bm{D}&=& \bm{D_s \otimes I_m } \text{ and } \bm{\Gamma }=\bm{\Gamma_s \otimes I_m}  \\
\bm{\Gamma_s}&=&\text{diag}(\cdots,\frac{1}{\tau_{ij}^2},\cdots) \in \R^{p \times p}
\end{eqnarray*} 
and  $\bm{D_s} \in \R^{p \times n}$ is a sparse matrix such that each row accommodates a $+1$ and $-1$, corresponding to $i \sim j$. We let $p$ denote the number of edges in the proximity network. Note that 
\begin{eqnarray*}
\bm{\beta' D' \Gamma D \beta }&=& \sum_{i \sim j} \frac{\| \bm{\beta_i - \beta_j}\|_2^2}{\tau_{ij}^2}.
\end{eqnarray*}
{ 
In light of the hierarchical representation, illustrated in equations (\ref{eq:t}, \ref{eq:hie}), the prior defined in  (\ref{eq:beta_prior}) can be viewed as an improper Gaussian mixture model; $\bm{\beta}$ is Gaussian given $\{\cdots,\tau_{ij}^2,\cdots\}$, and  the $\tau_{ij}^2$s come from a common ensemble. This prior favors spatial smoothness while allowing the amount of smoothness to be variable and adapt to the data.} As we will discuss in section \ref{sec:gibbs}, posterior samples of $\tau_{ij}^2$ tend to be smaller in smooth areas than in regions with discontinuities or outliers.

For each edge in the proximity network, and each corresponding row in $\bm{D_s}$, there is a unique pair of nodes $i$ and $j$ that are spatially ``nearby," i.e. $i \sim j$.   We found that considering the four horizontally and vertically nearby nodes as neighbors, for nodes that lie on a two dimensional regular lattice, allows us to efficiently estimate tuning functions without contamination from measurement noise or bias from oversmoothing. See section \ref{sec:opm} for an illustrative example. As for nodes that lie on an irregular grid, we compute the sample mean $\bm{\mu}$ and sample covariance $\bm{C}$  of the locations, and then whiten the location vectors $\bm{v_i}$; that is,  $\bm{v_{i,\text{whitened}}} = \bm{C^{-1/2} (v_{i} - \mu)}$. We found that connecting each node to its $k$-nearest-neighbors (within a maximum distance $r$) in the whitened space works well in practice. See section \ref{sec:motor} for an illustrative example with  $k=1$ and $r=5$. 

Extending the robust prior presented in equation (\ref{eq:beta_prior}), which is based on the simple local difference $\|\bm{\beta_i - \beta_j}\|_2$ for $i \sim j$, to a robust prior based on any generic $\| . \|_2$ measure of local roughness is easy; we only need to appropriately modify $\bm{D_s}$. For example, if $\bm{y_1,\cdots,y_n}$ are equidistant temporal samples, then the following robust prior
\begin{eqnarray*}
\bm{\beta}| \lambda,\sigma  &\sim& \prod_{i=1}^{n-2}\bigl ( \frac{\lambda}{2 \sigma} \bigr)^m \exp \Bigl( -\frac{\lambda}{\sigma} \Bigl \| \bm{ 2\beta_i - \beta_{i+1} - \beta_{i-1} }\Bigr\|_2 \Bigr) 
\end{eqnarray*} 
 reflects our a priori belief that $\bm{\beta_1,\cdots,\beta_n}$ are (approximately) piecewise linear \cite{KKBG09}. In this case, $\bm{D_s}$ is a tridiagonal matrix with 2 on the diagonal and -1 on the off diagonals. As another example, let the matrix $\bm{D_s}$ be equal to the discrete Laplacian operator; $[\bm{D_s}]_{ii}$ equals the number of edges attached to node $i$, and if $i \sim j$, then $[\bm{D_s}]_{ij}=-1$, otherwise its zero.  The discrete Laplacian operator (Laplacian matrix), which is an approximation to the continuous Laplace operator, is commonly used in the spatial smoothing literature to impose a roughness penalty \cite{Wahba90}.   Our robust prior based on the discrete Laplacian operator is as follows\begin{eqnarray*}
\bm{\beta} | \lambda,\sigma  &\sim& \prod_{i=1}^n\bigl ( \frac{\lambda}{2 \sigma} \bigr)^m \exp \Bigl( -\frac{\lambda}{\sigma} \Bigl \|  \sum_{j \sim i } (\bm{\beta_i - \beta_j}) \Bigr\|_2 \Bigr), 
\end{eqnarray*} 
which given the appropriate matrix $\bm{D_s}$ can easily be formulated in the hierarchical manner of equation \ref{eq:hie}. On regular grids, this prior is based only on the four (horizontal and vertical) neighbors but better approximations to the the continuous Laplace operator based on more neighbors is straightforward and within the scope of our scalable block Gibbs sampler presented in section  \ref{sec:gibbs}.

{ Finally note that the prior defined in (\ref{eq:beta_prior}) is not a proper probability distribution because it can not be normalized to one. However, in most cases the posterior distribution  will still be integrable even if we use such an improper prior \cite{Gelman03}. As we see later in section \ref{sec:gibbs}, all the conditional distributions needed for block-Gibbs sampling are proper.  Furthermore, the joint posterior inherits the unimodality in $\bm{\beta}$ and $\sigma$ given $\{ \nu_i \}_{i=1,\cdots,n}$  and $\lambda$  from the Bayesian Lasso \cite{PG08}, aiding in the mixing of the Markov chain sampling methods employed here. (See appendix \ref{sec:app}.)}

\begin{figure}
\hspace{-0.4cm}
        \centering
\includegraphics[width=16cm]{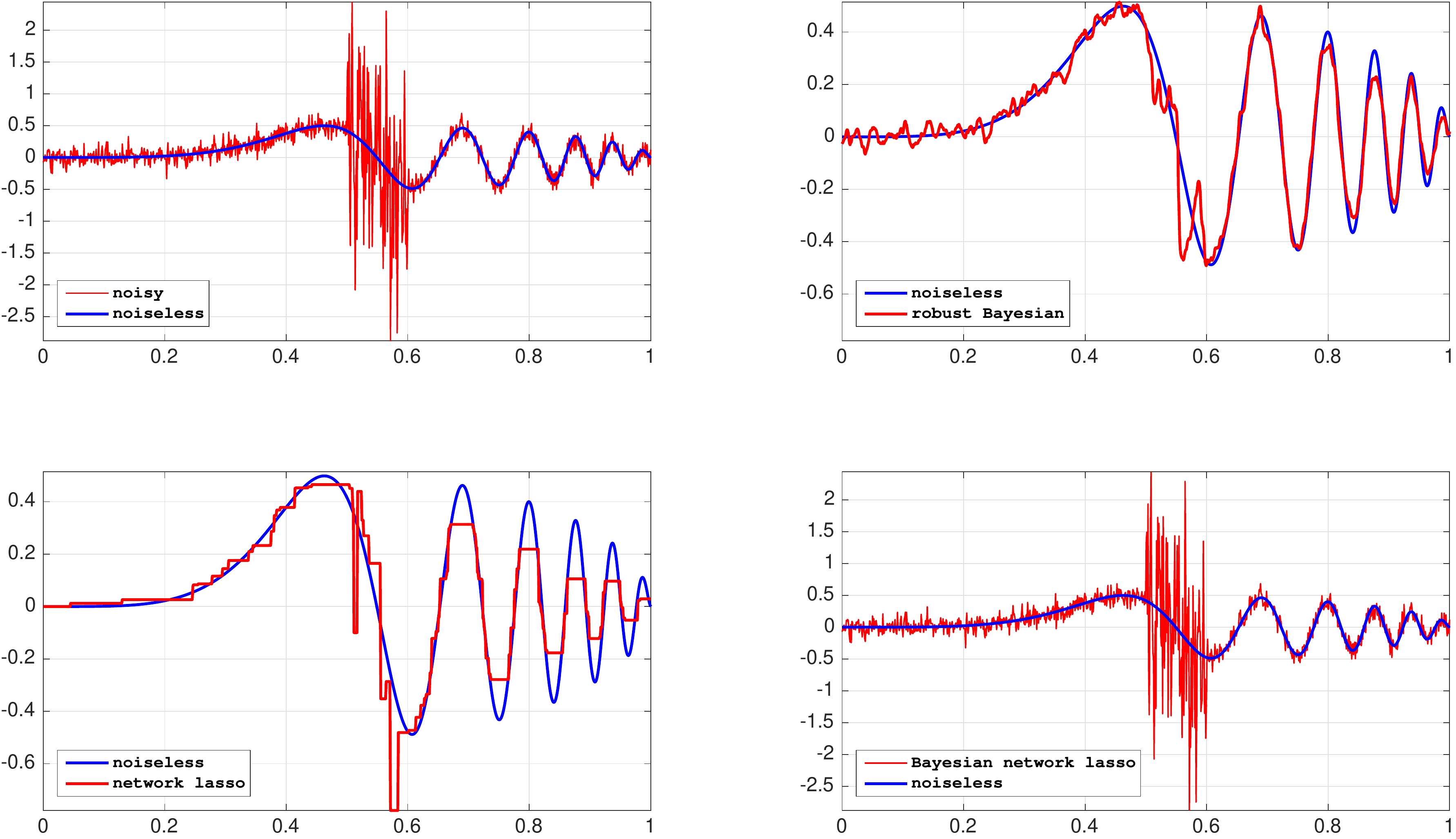} 
                \caption{{Heterogenous noise example. The Bayesian network lasso posterior mean estimate overfits in the region of higher observation noise. The robust Bayesian formulation is less prone to misidentifying heterogenous noise as signal. The network lasso tends to cluster high frequency variations into piecewise-constant estimates.}}
                \label{fig:heteroscedastic}
\end{figure}

\begin{figure}
\hspace{-0.4cm}
        \centering
                \includegraphics[width=16cm]{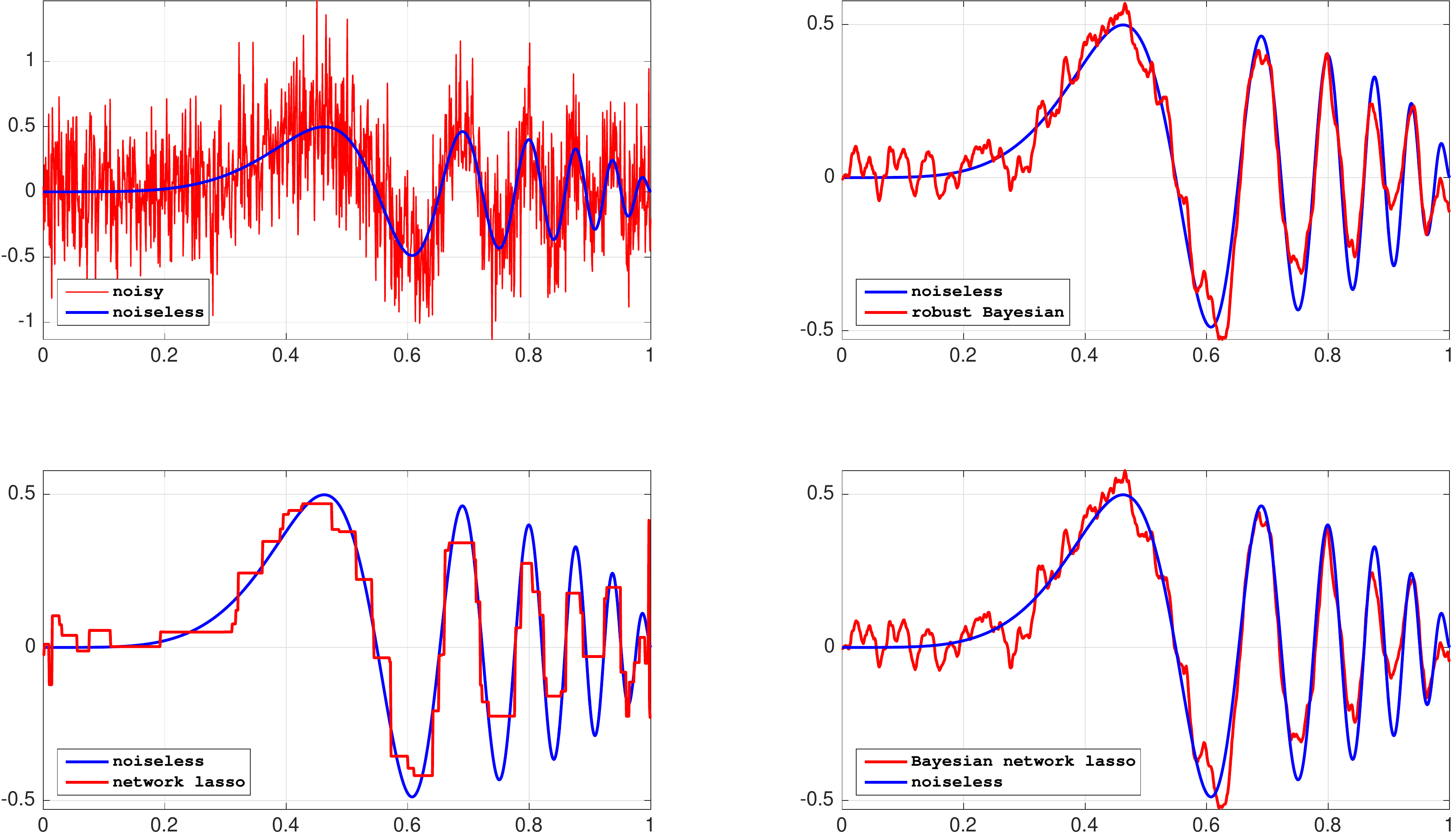}
                \caption{{Homogeneous noise example. The posterior means of Bayesian network lasso and our robust Bayesian are very similar. This is expected given the homogeneity of noise power. The network lasso suffers from the staircase effect, that is, the denoised signal is not
smooth but piecewise constant.}}
                \label{fig:homogeneous} 
\end{figure}

\begin{figure}
\hspace{-1.5cm}
        \centering
                \includegraphics[width=18cm]{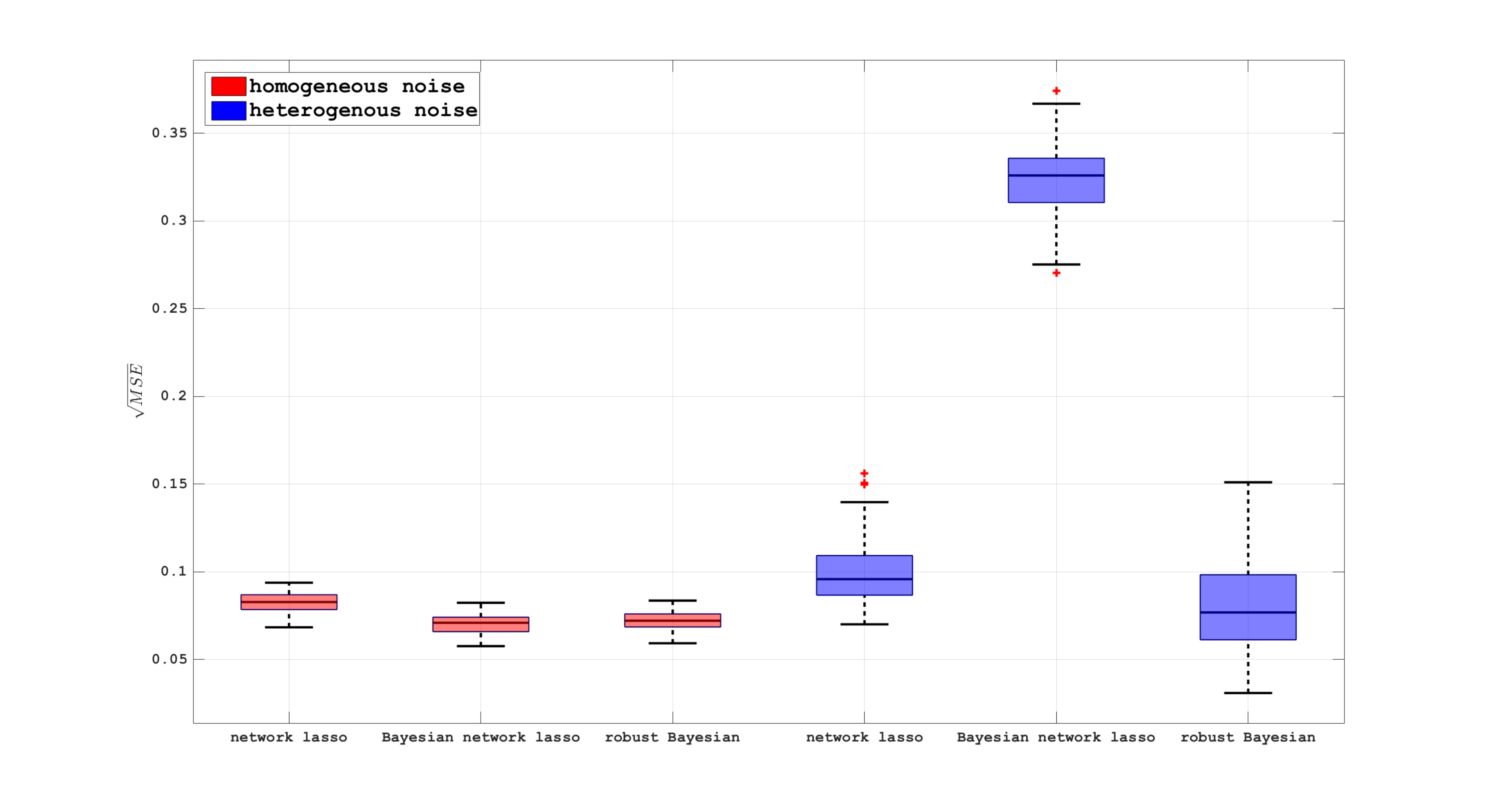}
                \caption{{Tukey boxplots comparing model $\sqrt{MSE}:=n^{-1/2}\| \bm{\beta} - \bm{\hat{\beta}}_{\text{model}} \|_2$ under homogeneous and heterogeneous noise. To make this comparison meaningful, signal-to-noise ratios are the same for both noise models. The boxplots are generated by simulating 100 replications of each model. For homogeneous noise, Bayesian network lasso and robust Bayesian perform similarly.  However, when noise is heterogeneous, Bayesian network lasso tends to overfit, as illustrated in figure \ref{fig:heteroscedastic}. In terms of $MSE$, network lasso is more robust to noise variations than its Bayesian counterpart but robust Bayesian performs slightly better.}}
                \label{fig:comparison}
\end{figure}

{
\subsection{Relationship to Network Lasso\label{sec:net_las}}
In related recent independent work, \cite{HLB15} present an algorithm based on the alternating direction method of multipliers \cite{ADMM11} to solve the network lasso convex optimization problem,
\begin{eqnarray}
\underset{\bm{\beta_i} \in \R^m \text{ for } i=1,\cdots,n}{\text{minimize}}  \hspace{1cm} \sum_{i=1}^n \|\bm{y_i - X_i \beta_i} \|_2^2 + \gamma \sum_{i \sim j} \|\bm{\beta_i -\beta_j} \|_2,\label{eq:netlasso}
\end{eqnarray}
in a distributed and scalable manner. The parameter $\gamma$  scales the edge penalty relative to the node objectives (and can be tuned using cross-validation). Similar to our formulation (section \ref{sec:ldn}), network lasso uses an edge cost that is a sum of norms of differences of the adjacent node variables, leading to a setting that allows for robust smoothing within clusters on graphs.  The optimization approach of \cite{HLB15} leads to fast computation but sacrifices the quantification of posterior uncertainty (which is in turn critical for closed-loop experimental design - e.g., deciding which neurons should be sampled more frequently to reduce posterior uncertainty) provided by the method proposed here.  A Bayesian version of the network lasso is a special case of our robust Bayesian formulation, by setting the variable variance parameters equal to one, that is $\nu_i^2=1$ for $i=1,\cdots,n$.  As we will see in the next example, heteroscedastic noise challenges the posterior mean estimate's robustness.


\subsection{Model Illustration}

In this section, we show that posterior means based on the prior of equation (\ref{eq:hpriors}) on $\{\nu_i\}_{i=1,\cdots,n}$ are robust to neuron-dependent noise variance. Our numerical experiments for heterogenous noise power show that a model with a homogeneous noise assumption will misinterpret noise as signal, depicted in figure \ref{fig:heteroscedastic}.  Comparisons with network lasso are presented as well. We postpone the details concerning the block-Gibbs sampler presented in this paper to  section \ref{sec:gibbs}.


The signal and heterogeneous noise models are as follows:
\begin{eqnarray*}
y_i &=& \beta_i + \epsilon_i, \hspace{2.4cm}
\text{ where } \hspace{1cm} \beta_i = \sqrt{\frac{i}{n}(1-\frac{i}{n})}\sin(11\pi \frac{i^4}{n^4}),\\
\epsilon_i &\sim& \mathcal{\N}(0,\sigma_i^2), \hspace{2.2cm} \text{ with     } \hspace{1.1cm} 
\sigma_i = \begin{cases}
    0.1       & \quad \text{if } \frac{i}{n} \in [0,0.5) \cup (0.6,1] \\
    1  & \quad \text{if } \frac{i}{n} \in [0.5,0.6]\\
  \end{cases}.
  \end{eqnarray*}
  The following hyperpriors were used for the posterior means of the robust Bayesian model:
  \begin{eqnarray*}
\sigma^2 &\sim&  \text{inverse-Gamma}(\kappa=0,\epsilon=0),\\
\lambda^2 &\sim&  \text{Gamma}(r=0.0001,\delta=0.001), \\
\nu_i^2 &\sim&   \text{inverse-Gamma}(\varkappa=3,\varepsilon=2).
\end{eqnarray*}
The hyperpriors of $\lambda^2$ and $\sigma^2$ are relatively flat. For $\nu_i^2$, we set the hyper parameters such that we have unit prior mean and prior variance. Bayesian network lasso is only different from the robust Bayesian formulation in that it assumes a constant noise variance, i.e. $\nu_i^2=1$ for $i=1,\cdots,n$.  

Bayesian network lasso and robust Bayesian posterior mean estimates are based on 10,000 consecutive iterations of the Gibbs sampler (after 5,000 burn-in iterations), discussed in section \ref{sec:gibbs}.  The network lasso estimate is the solution to the convex optimization problem equation (\ref{eq:netlasso}) where the tuning parameter $\gamma$ is set using 10-fold cross-validation. Note that the network lasso estimate corresponds to the mode of the posterior distribution of Bayesian network lasso conditioned on $\sigma$ and $\lambda$.

For the sake of comparison, we also present numerical results for a homogeneous noise model.  Here, the signal $\bm{\beta}$ is the same but the noise variance is $\sigma_i=0.33$ for $i=1,\cdots,n$. This particular choice of $\sigma_i$ was made to guarantee that the signal-to-noise ratio is equal to that of the heterogeneous noise model. As for the priors, they remain the same.  As expected, Bayesian network lasso and robust Bayesian posterior means are similar, depicted in figure \ref{fig:homogeneous}. 

Figures \ref{fig:heteroscedastic} and \ref{fig:homogeneous} illustrate that if the noise power is constant, robust Bayesian and Bayesian network lasso posterior means are similar. On the other hand, if noise power is not constant, robust Bayesian posterior mean detects the nonuniform noise power and adapts to it while Bayesian network lasso posterior mean will misinterpret noise as signal and overfit. Overall, network lasso estimates tend to over-smooth high frequency  variations into piecewise-constant estimates which is undesirable. Repeated simulations presented in figure \ref{fig:comparison} further confirm these observations.}


\section{Scalable Block Gibbs Sampling\label{sec:gibbs}}
We will now introduce some vector and matrix notation before we describe our Gibbs sampling approach to inference.  First, we introduce the following variables:
\begin{eqnarray}
\bm{\underline y_i} &:=& \frac{\bm{y_i}}{\nu_i}, \quad \bm{\underline X_i} := \frac{\bm{X_i}}{\nu_i}. \label{eq:u}
\end{eqnarray}
{We also let $\bm{\underline X} \in \R^{nd \times nm}$ stand for the rectangular blockwise-diagonal matrix  diag$\Bigl (\cdots,\bm{ \underline X_i},\cdots \Bigr)$.} Moreover, we let $\bm{\beta}$, $\bm{\underline y}$ and $\bm{\underline X' \underline y} $ stand for the column-wise concatenation (for $i=1,\cdots,n$) of $\bm{\beta_i}$, $\bm{\underline y_i}$, and $\bm{\underline X_i' \underline y_i}$, respectively. $\bm{\underline X' \underline X}$ is then the blockwise-diagonal matrix  diag$\Bigl (\cdots,\bm{\underline X_i' \underline X_i},\cdots \Bigr) \in \R^{nm \times nm}$. Finally, recall that  $p$ stands for the number of edges in the proximity network.

Our efficient Gibbs sampler and the full conditional distributions of $\bm{\beta}$, $\sigma^2$, $\{\nu_i^2 \}$, $\lambda$ and $\{ \tau_{ij}^2 \}$ can then be formulated as follows:

\textbf{Step 1.}  The local smoothing parameters $\{ \tau_{ij}\}_{i \sim j}$  are conditionally independent,  with { 
\begin{equation*}
\tau_{ij}^2 | \beta,\sigma^2,\lambda^2 \sim \bigl ( \frac{ 1}{ \tau_{ij}^2} \bigr)^{1/2}  \exp \Bigl ( - \frac{ \|\bm{\beta_i - \beta_j} \|^2}{2\sigma^2 \tau_{ij}^2}   -\frac{\lambda^2}{2} \tau_{ij}^2 \Bigr) \end{equation*}
}
\textbf{Step 2.}   The full conditional for $\bm{\beta}$ is multivariate normal with mean $\bm{P^{-1} \underline X'\underline y}$ and covariance $\sigma^2 \bm{P^{-1}}$, where $$\bm{P = \underline X'\underline X + D' \Gamma D}.$$

\textbf{Step 3.} $\sigma^2 \sim$ inverse-Gamma($\kappa'$,$\epsilon'$) with
\begin{equation*}
\kappa'=\kappa+ \frac{(pm+nd)}{2}, \quad\text{and} \quad \quad \epsilon' = \epsilon + \frac{1}{2} \| \bm{\underline y-\underline X\beta}\|^2 + \frac{1}{2}  \|\bm{\Gamma^{1/2} D \beta} \|^2.
\end{equation*}

\textbf{Step 4.} $\lambda^2 \sim $ Gamma($r'$,$\delta'$)  with
\begin{equation*}
r' = r + p(m+1)/2, \quad\text{and} \quad \quad \delta'=\delta+\frac{1}{2}\sum_{i \sim j} \tau_{ij}^2 .
\end{equation*}

\textbf{Step 5.} $\nu_i^2 \sim$ inverse-Gamma($\varkappa'$,$\varepsilon'$) with
\begin{equation*}
\varkappa'=\varkappa+ \frac{d}{2}, \quad\text{and} \quad \quad \varepsilon' = \varepsilon + \frac{1}{2\sigma^2} \| \bm{y_i-X_i\beta_i}\|^2 .
\end{equation*}

Note that in step 1, the conditional distribution can be rewritten as
\begin{eqnarray}
\frac{1}{\tau_{ij}^2} \Big | \bm{\beta},\sigma^2,\lambda \sim \text{inverse-Gaussian}(\mu',\lambda')  \label{eq:localsmooth}
\end{eqnarray}
with
\begin{equation*}
\mu'=\frac{\lambda \sigma}{\|\bm{\beta_i - \beta_j}\|_2},  \quad \quad \quad \lambda' = \lambda^2,
\end{equation*}
in the parametrization of the inverse-Gaussian density given by
\begin{equation*}
\text{inverse-Gaussian}(\mu',\lambda') \quad \sim \quad  f(x)=\sqrt{\frac{\lambda'}{2 \pi}} x^{-3/2} \exp \Bigl \{ - \frac{\lambda' (x-\mu')^2}{2 (\mu')^2 x}   \Bigr \}.
\end{equation*}
Moreover, the conditional expectation of $\frac{1}{\tau_{ij}^2}$ (using its inverse-Gaussian density in \ref{eq:localsmooth}) is equal to  $\frac{\lambda \sigma}{\|\bm{ \beta_i - \beta_j}\|_2} $. This makes the iterative Gibbs sampler above intuitively appealing; if the local difference  is significantly larger than typical noise (i.e., $\|\bm{\beta_i - \beta_j}\|_2  \gg \lambda \sigma$) then there is information in the difference, and therefore, minimal smoothing is applied in order to preserve that difference. On the other hand, if the local difference is small, this difference is likely to be due to noise, and therefore, local smoothing will reduce the noise. In other words, the robust Bayesian formulation presented in this paper functions as an adaptive smoother where  samples will be less smooth in regions marked with statistically significant local differences, and vice versa.

Furthermore in step 2, the conditional distribution of $\bm{\beta}$ depends on the observation $\bm{y}$ and the local smoothing parameters $\tau$. A large $1/\tau_{ij}^2$ causes the samples of $\bm{\beta_i}$ and $\bm{\beta_j}$ to be more similar to each other than their respective ML estimates $\bm{\beta_{i,\text{ml}}}$ and $\bm{\beta_{j,\text{ml}}}$ (where $\bm{\beta_{i,\text{ml}}} := \bm{(X_i'X_i)^{-1} X_i' y_i}$). In contrast, if  $1/\tau^2_{ij}$ is small, then the conditional samples of $\bm{\beta_i}$ and $\bm{\beta_j}$  typically revert to their respective ML estimates, plus block-independent noise.

Finally, although unnecessary in our approach, the fully Bayesian sampling of $\lambda$ in step 4 can be replaced with an empirical Bayes method. The difficulty in computing the marginal likelihood of $\lambda$, which requires a high-dimensional integration, can be avoided with the aid of the EM/Gibbs algorithm \cite{C01}. Specifically, iteration $k$ of the EM algorithm 
\begin{eqnarray*}
\lambda^{(k+1)} = \argmax_{\lambda} \E \Bigl [ \log p(\bm{ \beta}, \tau^2,\lambda | \bm{y}) \Big | \bm{y} ,\lambda^{(k)} \Bigr], 
\end{eqnarray*}
simplifies to
\begin{eqnarray}
\lambda^{(k+1)} &=& \sqrt{ \frac{p(m+1)}{\sum_{i \sim j} \E[\tau_{ij}^2| \bm{y},\lambda^{(k)} ] }} \label{eq:em},
\end{eqnarray}
which can be approximated by replacing conditional expectations with sample averages from step 1. The empirical Bayes approach gives consistent results with the fully Bayesian setting. The expectation of the conditional Gamma distribution of $\lambda^2$ in step 4:
\begin{eqnarray*}
\E[\lambda^2| \bm{y}, \tau] &=& \frac{2r+p(m+1)}{2\delta + \sum_{i \sim j} \tau_{ij}^2},
\end{eqnarray*}
is similar to the EM/Gibbs update (\ref{eq:em}). In our experience, both approaches give similar results on high dimensional data.

\subsection{Computational cost\label{sec:comp}}

The conditional independence of the local smoothing parameters $\{ \tau_{ij}\}_{i \sim j}$ given $\bm{\beta}$ and $\sigma$ amounts to a computational cost of sampling these variables that scales linearly with their size: $O(pm)$.  Similarly, the cost of sampling $\sigma^2$ given $\bm{\beta}$ and$\{ \tau_{ij}\}_{i \sim j}$ is due to computing $\sum_{i=1}^n\| \bm{y_i-X_i\beta_i}\|^2$, $\sum_{i=1}^n\| \bm{\underline y_i-\underline X_i\beta_i}\|^2$  and   $\|\bm{\Gamma^{1/2} D \beta }\|^2$ which are, respectively, $O(ndm)$, $O(ndm)$ and $O(pm)$, amounting to a total cost of $O((nd+p)m)$.

The conditional distribution of $\bm{\beta}$ given $\{ \tau_{ij}\}_{i \sim j}$ is multivariate Gaussian with mean $\bm{P^{-1} \underline X' \underline y}$ and covariance $\sigma^2 \bm{P^{-1}}$, whose computational feasibility rests primarily on the ability to solve the equation
\begin{eqnarray}
\bm{P w} &=& \bm{b}  \label{eq:lin}
\end{eqnarray}
as a function of the unknown vector $\bm{w}$, for $\bm{P} = \bm{\underline X' \underline X + D' \Gamma D}$. This is because if $\bm{\epsilon_1, \epsilon_2} \sim \N(0,\bm{I})$, then
\begin{eqnarray}
\bm{P^{-1} \underline X' \underline y }+ \sigma \bm{ P^{-1} }\Bigl [\bm{\underline X' \epsilon_1} + \bm{D' \Gamma^{1/2} \epsilon_2 }\Bigr]\label{eq:samp},
\end{eqnarray}
is a Gaussian random vector with mean $\bm{P^{-1} \underline X' \underline y}$ and covariance $\sigma^2\bm{ P^{-1}}$. { Similar approaches for the efficient realization of Gaussian fields based on  optimizing a randomly perturbed cost function (log-posterior)  were studied in \cite{HR91,H09,PaYu10,BSHL14,GMI14}.  In our case, the randomly perturbed cost function is
\begin{eqnarray*}
f_{\bm{\epsilon_1,\epsilon_2}} (\bm{\theta}) &:=& \bigl (\bm{D \theta} - \sigma \bm{\Gamma ^{-1/2} \epsilon_2}    \bigr)' \bm{\Gamma}   \bigl (\bm{D \theta} - \sigma \bm{\Gamma ^{-1/2} \epsilon_2}    \bigr) \\&+&  
\bigl (  \bm{\underline y } +\sigma \bm{\epsilon_1} - \bm{\underline X \theta}      \bigr)'\bigl (  \bm{\underline y } +\sigma \bm{\epsilon_1} - \bm{\underline X \theta}      \bigr),
\end{eqnarray*}
in which case it is easy to see that $\arg \max_{\bm{\theta}}f_{\bm{\epsilon_1,\epsilon_2}} (\bm{\theta})$ is given by equation (\ref{eq:samp}).
   }


Standard methods for computing $\bm{P^{-1}b}$ require cubic time and quadratic space, rendering them impractical for high-dimensional applications. A natural idea for reducing the computational burden involves exploiting the fact that  $\bm{P}$ is composed of a block-diagonal matrix   $\bm{\underline X' \underline X}$ and a sparse matrix $\bm{D'\Gamma D}$. { For instance, matrices based on discrete Laplace operators on regular grids lend themselves well to multigrid algorithms which have linear time complexity (see \cite{B77,GS89,PaYu10}  and section 19.6 of \cite{PRE92}).} Even standard methods for solving linear equations involving sparse matrices (as implemented, e.g., in MATLAB's $ \bm{P \backslash b} $ call) are quite efficient here, requiring sub-quadratic time \cite{RH05}. This sub-quadratic scaling requires that a good ordering is found to minimize fill-in during the forward sweep of the Gaussian elimination algorithm; code to find such a good ordering (via `approximate minimum degree' algorithms \cite{Davis06}) is built into the MATLAB call $\bm{P\backslash b}$  when $\bm{P}$ is represented as a sparse matrix. As we will see in section \ref{sec:opm},  exploiting these efficient linear algebra techniques permits sampling from a high dimensional ($> 10^6$) surface defined on a regular lattice in just a few seconds using MATLAB on a 2.53 GHz MacBook Pro.

\section{Motivating Neuroscience Applications\label{sec:results}}
\begin{figure}
\includegraphics[scale=0.75]{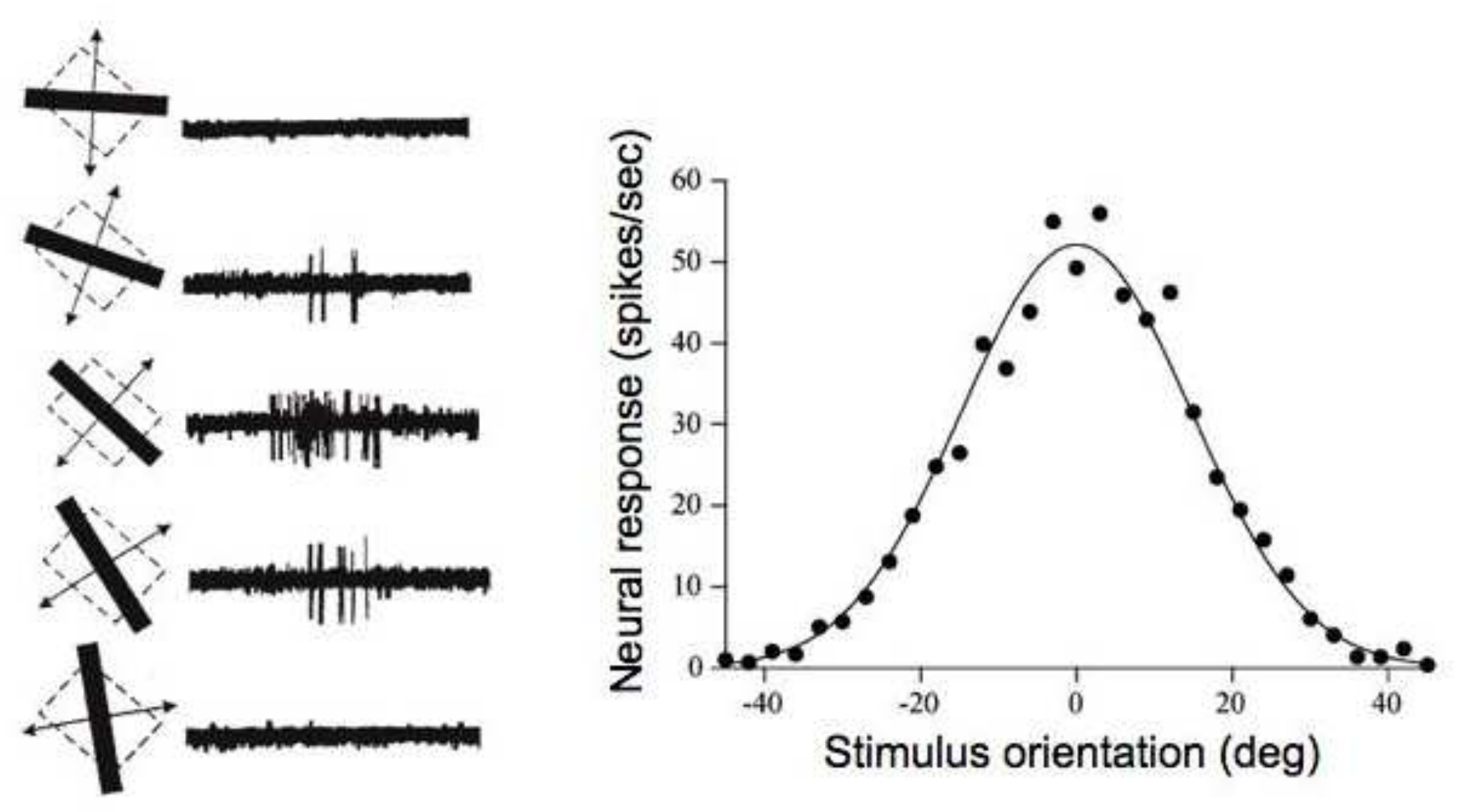}
\caption{Electrophysiological recordings from a single neuron in the primary visual cortex of a monkey. A moving bar of light  was projected onto the receptive field of the cell at different angles. In the diagrams on the left, the receptive field is shown as a dashed rectangle and the light source as a superimposed black bar. The angle of the dashed rectangle indicates the preferred orientation. For each bar (stimulus) orientation,  the neural response was recorded. The voltage traces in the  middle column show the electrophysiological recordings corresponding to the stimulus orientation of that row. Note that the neural response depends on the stimulus orientation; it increases as the bar and the preferred orientation become more aligned. Clearly, the bar orientation of the middle row evoked the largest number of action potentials. The graph on the right shows average number of action potentials per second (neural response) versus the angle of the bar. This graph indicates how the neural response depends on the orientation of the light bar. The data have been fit by a Gaussian function. (Data is from \cite{HW68,HDB74} and figures are adapted from \cite{WAND95,DA01}.) } 
\label{fig:opm_data}
\end{figure}
\begin{figure}
\hspace{0cm}
\includegraphics[width=0.95\textwidth]{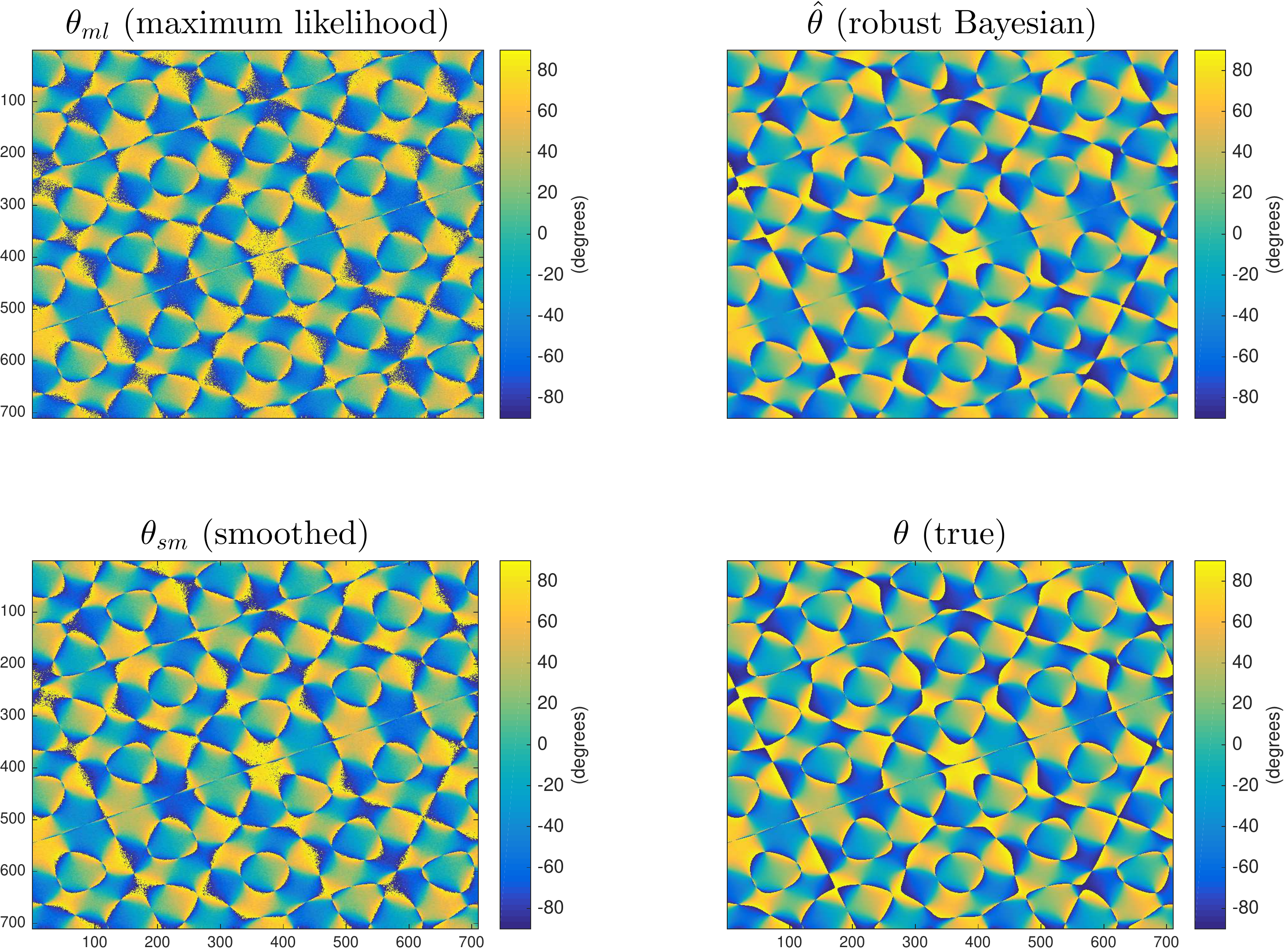}
\caption{ Analysis of a synthetic orientation tuning map. $\theta$ is a synthetic $710 \times 710$ orientation preference map (see section 2.4 of the Supplement of \cite{K10} for details).  Each pixel is a neuron, and $\theta_i \in (-90^\circ,+90^\circ]$ (the preferred orientation  of neuron $i$) is given by $\arctan \bigl (\beta_{2,i} /\beta_{1,i}  \bigr)$. Likewise, the robust Bayesian $\hat \theta$, smoothed $\theta_{sm}$, and maximum-likelihood $\theta_{ml}$ estimates of preferred orientations are inverse trigonometric functions of $\bm{\hat \beta}$, $\bm{\beta_{sm}}$, and $\bm{\beta_{ml}}$, respectively. The Bayesian estimate $\hat \theta$ of preferred orientations is less noisy than $\theta_{ml}$  and more robust than $\theta_{sm}$; see also Fig.~\ref{fig:r_theta} for a  zoomed-in view. The $\bm{\hat \beta}$ estimate of posterior expectations is based on 10000 consecutive iterations of the Gibbs sampler (after 500 burn-in iterations). 
}                
\label{fig:orientation_preference}
\end{figure}

\begin{figure}
\hspace{0cm}
                \includegraphics[width=1\textwidth]{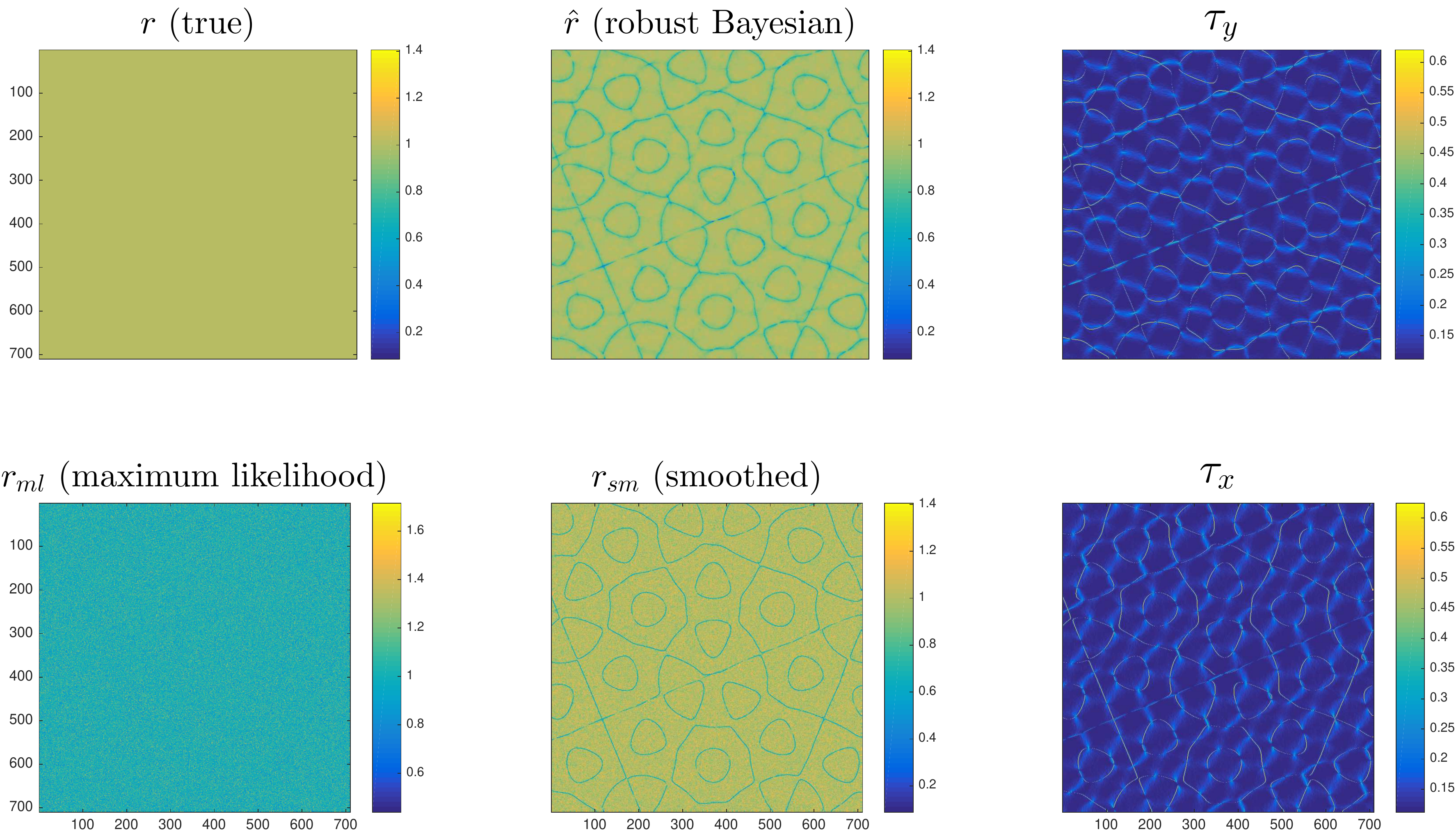}
                \caption{True tuning strengths $\{r_i\}$, the estimated tuning strengths $\{ r_{i,ml},r_{i,sm},\hat r_i \}$, and posterior means of local smoothing parameters $\{\tau_{ij}\}_{i \sim j}$. Each pixel is a neuron, and its (estimated) tuning strength is given by the length of its (estimated) $\bm{\beta_i}$, e.g.  $r_i = \| \bm{ \beta_i }\|_2$, $\hat r_i = \| \bm{\hat \beta_{i} }\|_2$, etc. The proximity network is a $710 \times 710$ regular grid with edges between a node and its four (horizontal and vertical) neighbors. The local smoothing parameters defined on edges among vertical and horizontal edges are designated by $\{\tau_y\}$ and $\{\tau_x\}$, respectively. The $r_{sm}$ (smoothed) and $\hat r$ (robust Bayesian)  tuning strength maps underestimate the true value at points where posterior means of local smoothing parameters $\{\tau_x,\tau_y \}$ take significant values. These points correspond to   sharp breaks in the orientation preference map $\theta$ (as illustrated in figure \ref{fig:orientation_preference}) where local averaging of significantly differently oriented tuning functions leads to a downward bias in estimated tuning strengths.}

\label{fig:r_tau}
\end{figure}

\begin{figure}
\hspace{0cm}
                \includegraphics[width=1\textwidth]{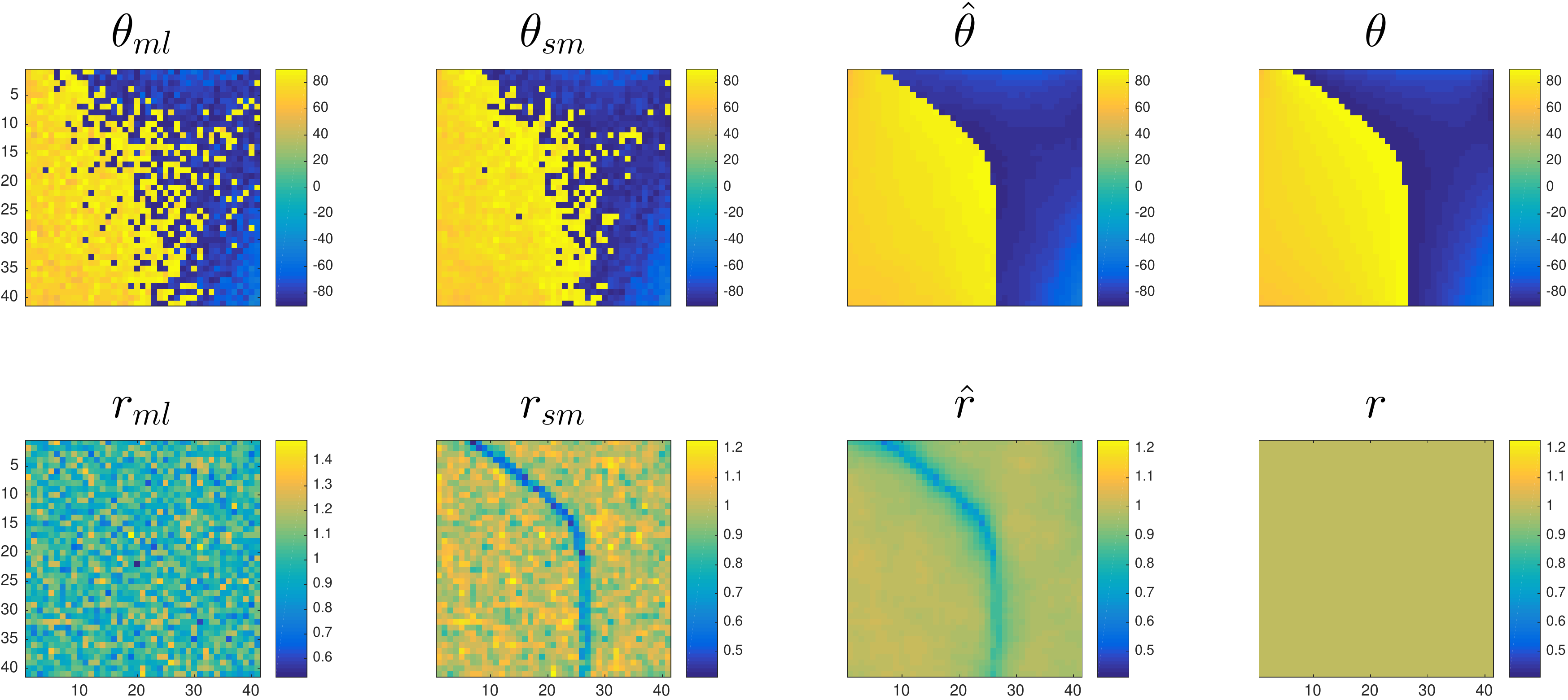}
                \caption{ A $40 \times 40$ zoomed-in view of preferred orientations $\{\theta_i\}$ and tuning strengths $\{r_i\}$, and their estimates. (The center of this map is pixel (241, 60) in figure \ref{fig:orientation_preference} and figure \ref{fig:r_tau}.)  The smoothed $r_{sm}$   tuning strength map underestimates the true tuning strength at sharp breaks in the orientation preference map $\theta$.  This bias is less severe for the Bayesian estimate $\hat r$ because the robust prior applies less local smoothing at sharp breaks (as illustrated in figure \ref{fig:r_tau}).  Similarly, $\hat \theta$ provides much more accurate angular estimates than $\theta_{sm}$.                  
            }               
\label{fig:r_theta}
\end{figure}

\begin{figure}
\hspace{0cm}
                \includegraphics[width=1\textwidth]{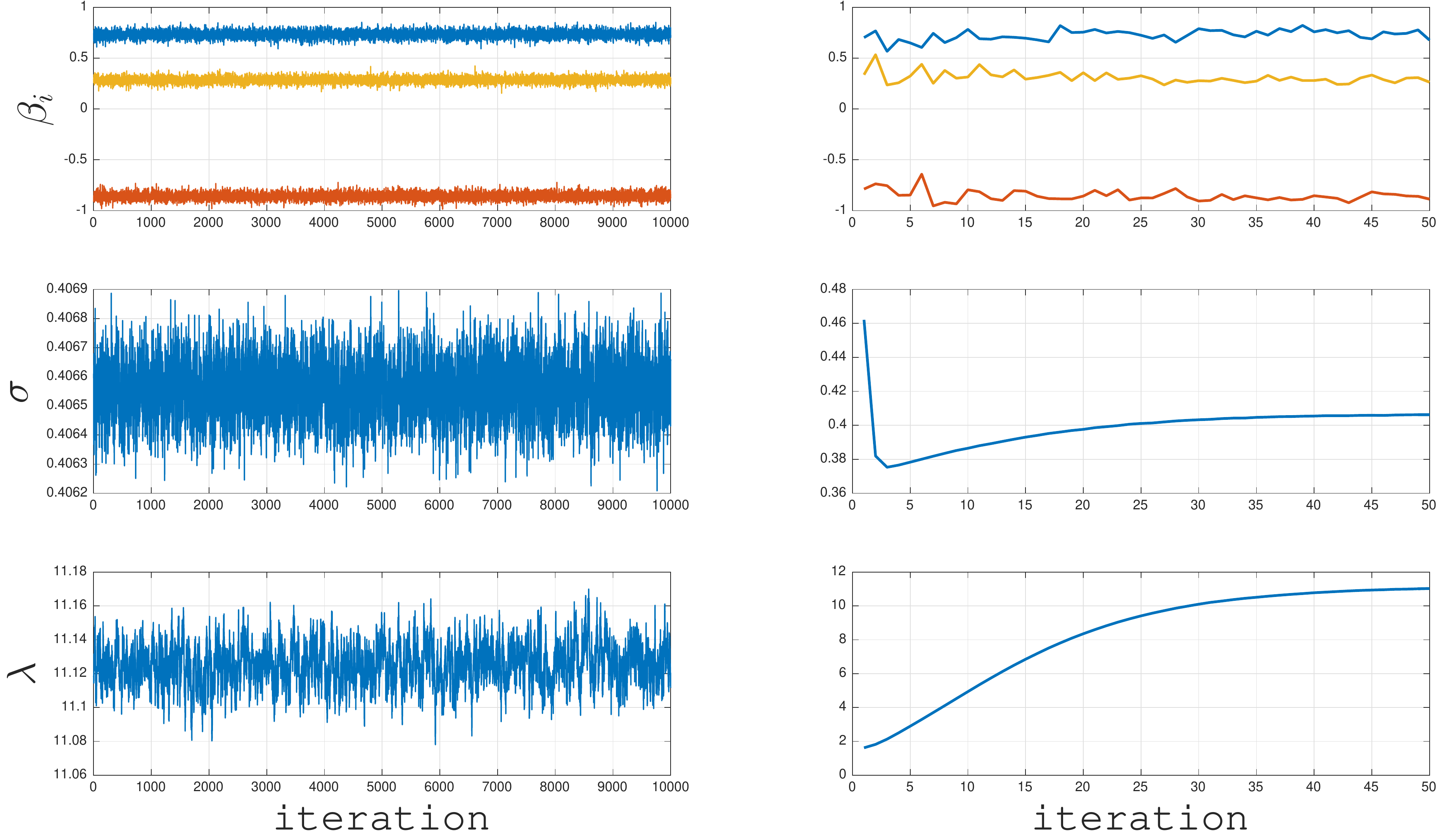}
                \caption{ { The sample path of 3 randomly selected pixels (top), $\sigma$ (middle), and $\lambda$ (bottom). The last 10000 (after 500 burn-in) samples (left) and the first 50 samples (right).}
                }
\label{fig:samples}
\end{figure}


\begin{figure}
\hspace{0cm}
                \includegraphics[width=1\textwidth]{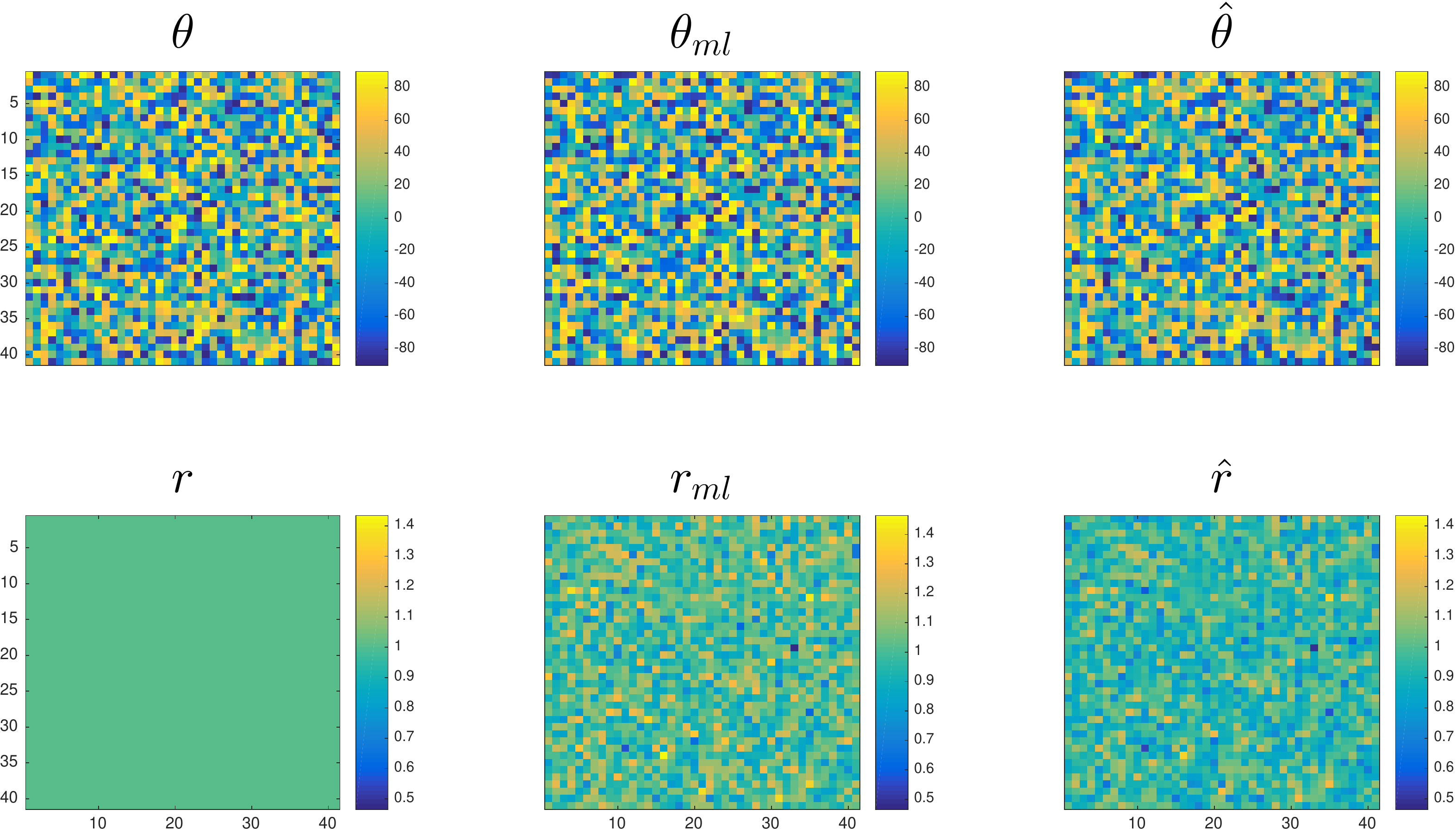}
                \caption{ { A $40 \times 40$ zoomed-in view of the $710 \times 710$ (not shown) randomly arranged preferred orientations $\{\theta_i\}$ and tuning strengths $\{r_i\}$, and their estimates. The orientation at each pixel was randomly drawn from a uniform distribution on $(-90^\circ,+90^\circ]$. Since the preferred orientations lack spatial organization,  the Bayesian estimate $\hat \theta$ of preferred orientations reverts to its respective $\theta_{ml}$.  The posterior estimates are based on 10000 consecutive iterations of the Gibbs sampler (after 500 burn-in iterations).}}               
\label{fig:orientation_preference_mouse}
\end{figure}

Here we will discuss the application of our robust Bayesian analysis approach towards the analysis of both synthetic and real neural tuning maps. In both cases, our new algorithm permits the robust estimation of neural tuning with higher fidelity and less data than alternative approaches.

\subsection{Synthetic data}
\subsubsection{Estimating Orientation Preference Maps\label{sec:opm}}

We will first apply our algorithm to synthetic data modeled after experiments where an animal is presented with a visual stimulus and the neural activity in primary visual cortex (also known as V1) is simultaneously recorded. V1 is the first stage of cortical visual information processing and includes neurons that selectively respond to sinusoidal grating stimuli that are oriented in specific directions \cite{HW62}. Neurons with such response properties are called \textit{simple cells}. See figure \ref{fig:opm_data} for an illustrative example of the recorded neural activity while a bar of light is moved at different angles  \cite{HW68,HDB74,WAND95,DA01}. As can be seen the figure, action potential firing of the simple cell depends on the angle of orientation of the stimulus. 

To capture the essential characteristics of simple cells in the visual cortex, we will use the following model. The response of cell $i \in \{1,2,\cdots,n \}$ to a grating stimulus with orientation $\phi_{\ell}$ depends on the preferred orientation  $\theta_i \in (-90^\circ,+90^\circ]$, and the tuning strength $r_i \in \R^{+}$ of that cell. The number of cells is $n$, and the number of trials (with differently oriented stimuli)  is $d$. Formally speaking, in the simplest linear model, during the $\ell$th trial, the noisy measurement $y_{i,\ell} \in \R$ at neuron $i$ in response to a stimulus with orientation $\phi_{\ell}$ can be written as \cite{S98,MGWKB10,MGWKB11}
\begin{eqnarray*}
y_{i,\ell} \big | \bm{\beta_i,x_{\ell}},\sigma^2 &\sim& \N(\bm{\beta_i' x_{\ell}} ,\sigma^2 ), \quad i=1,\cdots, n \quad \text{and} \quad \ell=1,\cdots,d,
\end{eqnarray*}
where {$\bm{\beta_i}:= r_i[\cos\theta_i \ \sin\theta_i]'$} is related to $\theta_i$ (preferred orientation)  and $r_i$ (tuning strength) as follows
$$
\theta_i:=\arctan \Bigl[ \frac{\beta_{2,i}}{\beta_{1,i}}\Bigr],
\quad \quad
r_i := \sqrt{\beta_{2,i}^2+\beta_{1,i}^2},
$$
and {$\bm{x_{\ell}}=[\cos\phi_{\ell} \ \sin\phi_{\ell}]'$} stands for the grating stimulus with orientation $\phi_{\ell}$. Writing the stimulus set $\{ \bm{x_{\ell}} \}_{\ell=1,\cdots,d}$ in matrix notation
\begin{eqnarray*}
\bm{X_{\o}} &:=& \begin{bmatrix}
        \vdots  \\
     \bm{ x_{\ell}'} \\
       \vdots  \\
     \end{bmatrix}_{d \times 2},
\end{eqnarray*}
allows us to compactly rewrite the neural response $\bm{y_i }\in \R^d$  as
\begin{equation}
\bm{y_i} \big | \bm{X_{\o}, \beta_i}, \sigma^2 \sim    \N (\bm{X_{\o} \beta_i}, \sigma^2 \bm{I}) \quad i=1,\cdots, n \label{eq:1}.
\end{equation}
Note that all neurons respond to the same particular grating stimulus, namely $\bm{X_{\o}}$, though due to different preferred orientations, not
all neurons respond similarly. 

In this example, the noise variances are set to be equal, that is $\nu_i=1$ for $i=1,\cdots,n$. As for the Gibbs sampler, we skip step 5, and substitute $\nu_i=1$ in all other steps. In the next section, we present a real data example, where  $\{\nu_i\}$ is estimated using step 5 of our Gibbs sampler. 


Drawing conclusions regarding the cortical circuitry underlying orientation maps, their formation during visual development, and across evolution, has recently been the subject of numerous studies \cite{SKLW07,RLW09,K10,KKS12}. For instance, \cite{K10} argued that evolutionary history (instead of ecological or developmental constraints) underlies the formation of qualitatively similar pinwheel distributions observed in the visual cortex of disparate mammalian taxa. Consequently, the estimation of orientation maps without contamination from measurement noise or bias from overs-smoothing will help to clarify important questions about evolution and information processing in the visual cortex.

We therefore generated synthetic tuning maps by extracting the phase of superpositions of complex plane waves (see section 2.4 of the Supplement of \cite{K10} for details). In our simulations, for clarity we assume $\bm{\beta_i}=(\cos \theta_i, \sin \theta_i)'$, and therefore $r_i=1$, which means tuning strengths are constant across all neurons. The top left panels of figure \ref{fig:orientation_preference} and figure \ref{fig:r_tau} show the angular components  $ \{ \theta_i\}$  and tuning strengths $r_i=1$ of the resulting map. It is well known that in some species the preferred orientations $ \{ \theta_i\}$ are arranged around singularities, called pinwheel centers  \cite{OHKI05,OhkiReid06}. Around each singularity, the preferred orientations $ \{ \theta_i\}$ are circularly arranged, resembling a spiral staircase. If we closely examine the top left panel of figure \ref{fig:orientation_preference}, it is evident that around pinwheel centers the preferred orientations $ \{ \theta_i\}$ are descending, either clockwise or counterclockwise from 
$-90^\circ$ to $+90^\circ$. Experimentally measured maps obtained from cats, primates, \cite{K10} and our synthetically generated data all share this important feature.


{ We simulated the neural responses of each cell to twenty differently oriented grating stimuli  by sampling responses according to equation (\ref{eq:1}) with $\sigma=0.4$. The orientations $\phi_{\ell}$ (for $\ell=1,\cdots,20$) were randomly and uniformly sampled from $(-90^\circ,+90^\circ]$.} Our main objective is to estimate (from neural responses $\{\bm{y_i }\}$ and stimuli $\bm{X_{\o}}$) the preferred orientations $\{ \theta_i \}$ and tuning strengths $\{ r_i \}$.   Ordinary linear regression yields maximum-likelihood estimates 
\begin{eqnarray}
\bm{\beta_{i,\text{ml}}} &=& \bm{(X_{\o}'X_{\o})^{-1}X_{\o}' y_i }\label{eq:ml}\\
\theta_{i,\text{ml}} &=& \arctan \Bigl( \frac{   \beta_{2,i,\text{ml}}} {\beta_{1,i,\text{ml}} } \Bigr),\nonumber \\
r_{i,\text{ml}} &=& \| \bm{\beta_{i,\text{ml}}} \|_2. \nonumber
\end{eqnarray}
The maximum likelihood estimates $\theta_{i,\text{ml}}$ and $r_{i,\text{ml}}$ are depicted in figure \ref{fig:orientation_preference}, \ref{fig:r_tau} and \ref{fig:r_theta}. The fine structure around pinwheel centers and the border between clustered preferred orientations is disordered.

We also computed the smoothed estimate $\bm{\beta_{sm}}$ based the following smoothing prior
\begin{eqnarray*}
p(\bm{\beta} | \gamma) &\propto& \exp \Bigl ( -\frac{\gamma}{2} \sum_{i \sim j }\| \bm{\beta_i - \beta_j} \|_2^2  \Bigr)\\
&\propto& \exp \Bigl ( -\frac{\gamma}{2} \bm{\beta' D' D \beta}  \Bigr),
\end{eqnarray*}
and the likelihood in (\ref{eq:1})
\begin{eqnarray*}
p(\bm{y} | \bm{\beta}) &\propto& \exp \Bigl ( -\frac{1}{2\sigma^2} \| \bm{y - X\beta} \|_2^2  \Bigr),
\end{eqnarray*}
leading to  following posterior expectation of $\bm{\beta}$:
\begin{eqnarray}
\bm{\beta_{sm}}(\gamma) := (\bm{X'X} + \gamma \bm{D' D})^{-1} \bm{X'y} \label{eq:sm},
\end{eqnarray}
where $\bm{X'X = I_{n \times n}  \otimes X_{\o}'X_{\o}  }$ and $\bm{X'y}=(\cdots, \bm{X_{\o}'y_i},\cdots)$. The smoothed estimate $\bm{\beta_{sm}}$ is based on a Gaussian prior that penalizes large local differences quadratically.  (In contrast, the robust prior defined in equation \ref{eq:beta_prior} penalizes large differences linearly.) The amount of smoothing is dictated by $\gamma$; large values of $\gamma$ lead to over-smoothing and small values of $\gamma$ lead to under-smoothing. In this example, the true $\beta$ is known; therefore, for the sake of finding the best achievable smoothing performance, { we selected $\gamma=2.15$ }(using a grid search)  which minimizes $\|\bm{\beta_{sm}}(\gamma)-\bm{\beta} \|_2$. 

The proximity network that we used in this example was defined using the edges between every node and its four nearest (horizontal and vertical) neighbors. The smoothed estimates $\theta_{i,\text{sm}}:= \arctan \Bigl ( \frac{  \hat  \beta_{2,i,sm}} {\hat \beta_{1,i,sm} } \Bigr)$ and $r_{i,\text{sm}}:=\|\bm{\beta_{i,sm}}\|_2$ are depicted in figures \ref{fig:orientation_preference}, \ref{fig:r_tau} and \ref{fig:r_theta}.  In spite of the observation that  $\theta_{sm}$ is less noisy than $\theta_{ml}$, there are still areas where the fine structure around pinwheel centers and the border between clustered preferred orientations is disordered. 

Figure \ref{fig:r_tau} shows that $r_{sm}$ is typically close to the true value of one, except for in neurons that lie at the border between regions with different orientation preferences. This is due to the fact that at regions that mark the border, tuning functions (and their noisy observations) point at significantly different directions, and therefore, local averaging decreases the length of the average value. On the other hand, in smooth regions where vectors are pointing in roughly the same direction, local averaging preserves vector length. 

The ability of our method to recover orientation preference maps from noisy recordings is shown in figures \ref{fig:orientation_preference}, \ref{fig:r_tau} and \ref{fig:r_theta}. To use the Bayesian formulation of equation \ref{eq:beta_prior}, we substituted a fixed $\bm{X_{\o}}$ for all $\bm{X_i}$. {For $\lambda^2$, a Gamma($r=1,\delta=1$) was used based on the understanding that a priori $\frac{1}{p}\sum_{i \sim j}\|\beta_i -\beta_j\|_2 $  should be $O(1)$. As for $\sigma^2$, the improper inverse-Gamma($\kappa'=0$,$\epsilon'=0$), i.e. $\pi(\sigma^2) \propto 1/\sigma^2$, was used. $\{\bm{ \hat \beta_i}\}$, namely the posterior expectation of $\{\bm{\beta_i}\}$, is based on 10000 samples from our efficient Gibbs sampler (after 500 burn-in iterations). The estimates $\hat \sigma=0.4066 \pm  0.0001$  and $\hat \lambda=11.13 \pm 0.01 $ (i.e., the mean $\pm$ standard deviation) are based on the 10000 samples.} The following estimates of the preferred orientations and tuning strengths
\begin{eqnarray*}
\hat \theta_i &:=& \arctan \Bigl ( \frac{   \hat \beta_{2,i}} {\hat \beta_{1,i} } \Bigr),\\
\hat r_i &:=& \| \bm{\hat \beta_i} \|_2,
\end{eqnarray*}
are depicted in  figures \ref{fig:orientation_preference} and \ref{fig:r_tau}. The posterior mean estimates of $\tau_x$ and $\tau_y$ (depicted in figure \ref{fig:r_tau}) tend to be larger for neurons on the border of regions with similar preferred orientations $\{ \theta_i \}$ (and less so around pinwheel centers), leading to minimal local smoothing for those pixels. Figure \ref{fig:r_tau} shows that the Bayesian estimate $\hat r$  (like $r_{sm}$) underestimates the tuning strength for points that mark the border between different orientation preferences.  In comparison to $r_{sm}$, as illustrated in the zoomed-in maps of figure \ref{fig:r_theta}, this problem is less severe for the Bayesian estimate $\hat r$ because of the robust prior that decreases the strength of local averaging by increasing the local smoothing parameters $\{ \tau_{ij} \}$ in regions marked with discontinuities.

As we can see in figure \ref{fig:r_theta}, the sharp border between similar orientation preferences is not over-smoothed  while the noise among nearby neurons with similar orientation preferences is reduced.  As a consequence of robustness, information is shared less among cells that lie at the border, but for cells that lie inside regions with smoothly varying preferred orientation, local smoothing is stronger. Moreover, in this example the chain appears to mix well (see figure \ref{fig:samples}), and  the Gibbs sampler is computationally efficient, requiring just a few seconds on a laptop (per iteration) to sample a surface described by $> 10^6$ parameters. 

{Finally, let us add that it is well known that the semiregular, smoothly varying arrangement (with local discontinuities) of orientation preference maps is not a general feature of cortical architecture \cite{VHCNT05}. In fact, numerous electrophysiological and imaging studies \cite{TB76,MB79,MGI88,GSL99}  have found that orientation selective neurons in the visual cortex of many rodents are randomly arranged. A question that arises is whether the model would over-smooth if the neurons are not arranged smoothly in terms of their maps. In order to answer  this question,  we generated a randomly arranged orientation preference map, and applied our algorithm to the simulated neural activity in response to the  same grating stimuli $\bm{X_{\o}}$ used above. We also used the same noise variance ($\sigma=0.4$) and the same priors for $\lambda,\sigma$ and $\{\bm{\beta} \}_{i=1,\cdots,n}$.  Results are depicted in figure \ref{fig:orientation_preference_mouse}. Since the preferred orientations lack spatial organization, the Bayesian estimate $\hat \theta$ of preferred orientations reverts to its respective $\theta_{ml}$. }
\subsection{Real data}
\subsubsection{Phasic tuning in motor neurons\label{sec:motor}}

\begin{figure}
\includegraphics[width=1\textwidth]{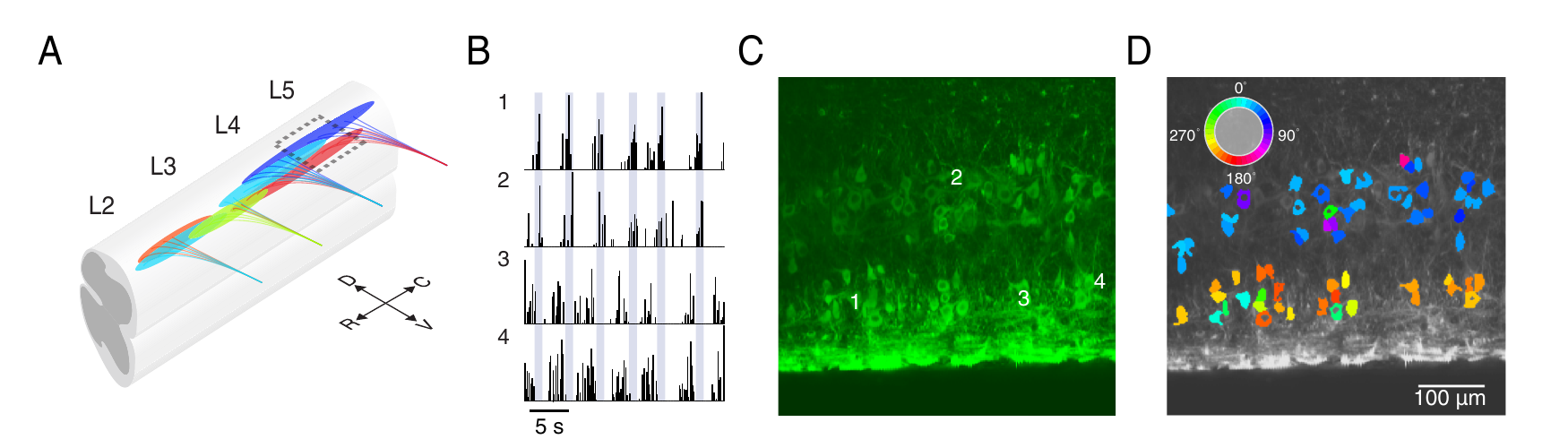}
\caption{Isolated spinal cord imaging preparation. (a) Schematic of isolated spinal cord imaging preparation. (b) Activity inferred from fluorescence measurements obtained from four motor neurons. Height of black bars indicates intensity of neuronal activity at each time point. Vertical blue bars indicate the onset of each locomotor cycle (i.e. $0^{\circ}$). (c) Example fluorescent imaging field with the position of the four neurons shown in (b) indicated. (d) Each motor neuron shown in (c) is represented in color (legend in inset) corresponding to its estimated tuning value.}\label{fig:spinal-data} 
\end{figure}

We next tested the method's performance on real neural imaging data obtained from an isolated mouse spinal cord preparation (schematized in figure \ref{fig:spinal-data}A). In these data, the fluorescent activity sensor GCaMP3 was expressed in motor neurons that innervate leg muscles. After application of a cocktail of rhythmogenic drugs, all motor neurons in the preparation fire in a periodic bursting pattern mimicking that seen during walking \cite{MPPJM15}. Under these conditions, we acquired sequences of fluorescent images and then applied a model-based constrained deconvolution algorithm to infer the timing of neuronal firing underlying each fluorescent activity time series extracted from the pixels corresponding to individual neurons \cite{Eftychios_2014}.

\begin{figure}
        \centering
         \begin{subfigure}[b]{0.3\textwidth}
                \includegraphics[width=\textwidth]{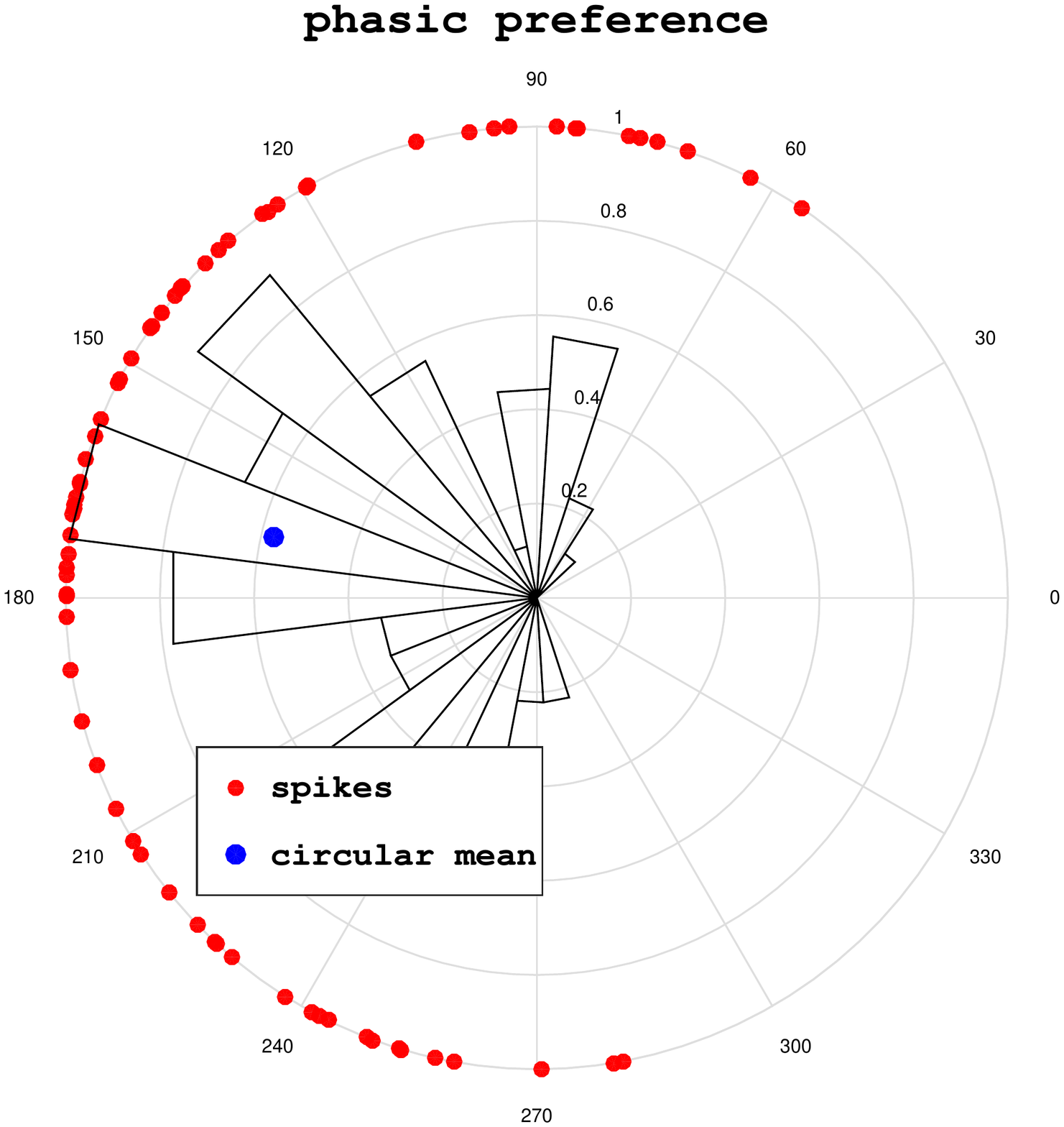}
                \caption{ Phasic preference of one cell}
                \label{fig:raw_single}
        \end{subfigure}%
        ~ 
        \begin{subfigure}[b]{0.3\textwidth}
                \includegraphics[width=\textwidth]{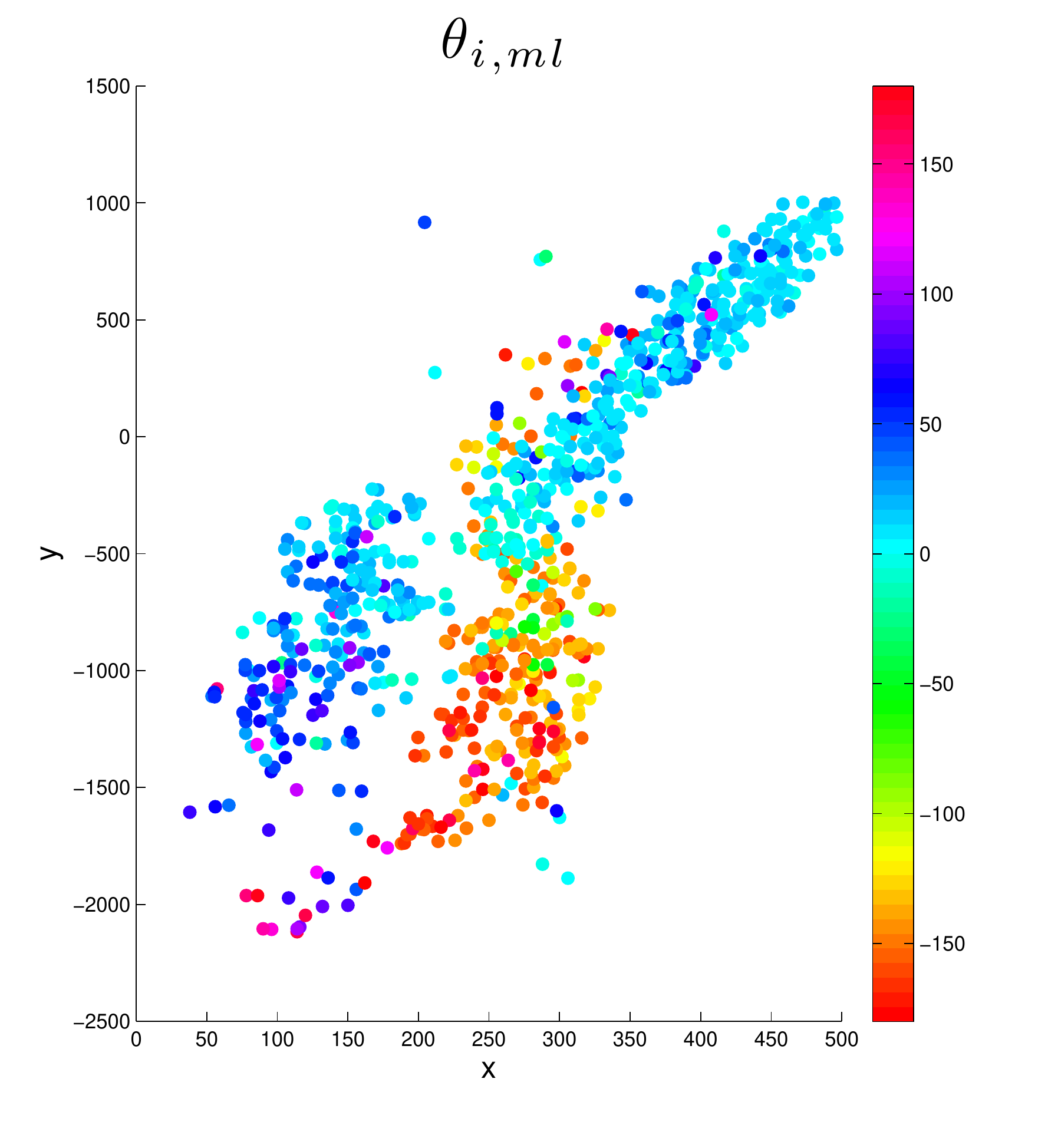}
                \caption{ Preferred phases of $n=854$ cells.}
                \label{fig:raw_pp}
        \end{subfigure}%
        ~ 
        \begin{subfigure}[b]{0.3\textwidth}
                \includegraphics[width=\textwidth]{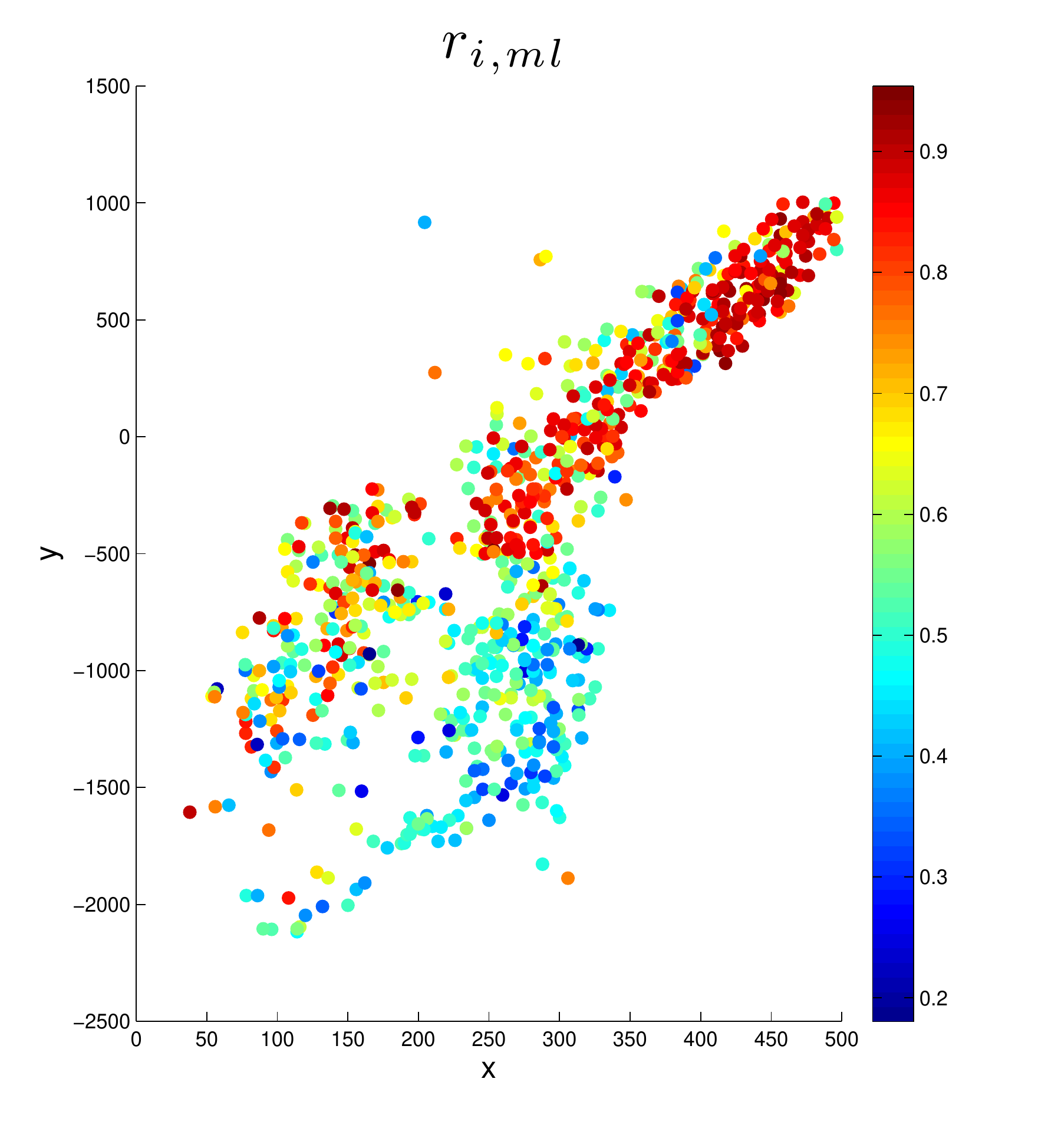}
                \caption{ Tuning strengths of $n=854$ cells.}
                \label{fig:raw_ts}
        \end{subfigure}
                \caption{(a) Noisy observations of phases at which this cell has fired. The phase of each red dot on the unit circle is a phase at which this cell has fired, and the angular histogram depicts its distribution. The blue dot is the circular mean of all red dots, and its phase and length are the ML estimates of the preferred phase and tuning strength, respectively. (b,c)  The three-dimensional spatial cell position is projected into the two-dimensional x-y plane. Each dot indicates one cell; each cell is color coded with the phase $\theta_{i,\text{ml}}$ or tuning strength $r_{i,\text{ml}}$.  Preferred phases (and tuning strengths) tend to be similar among nearby cells, but not all nearby cells have similar preferred phases (and tuning strengths).}\label{fig:raw}
\end{figure}

Each mouse leg is controlled by $\sim50$ different muscles, each of which is innervated by motor neurons that fire in distinct patterns during locomotor behavior \cite{Krouchev_2006,Akay_2014}. Furthermore, all motor neurons that share common muscle targets are spatially clustered together into ``pools" within the spinal cord \cite{romanes1964motor}. Therefore, during the locomotor-like network state monitored in these data, different spatially-distinct groups of motor neurons are recruited to fire at each moment in time (figure \ref{fig:spinal-data}B-D). When the activity of each motor neuron is summarized as a single mean phase tuning value (representing the average phase angle of the $\sim70$ firing events detected per neuron, as seen in figure \ref{fig:raw_single}), a clear spatial map can be derived (figure \ref{fig:spinal-data}D). Such maps appear smooth within pools, and sharply discontinuous between pools.

While phase tuning can be reliably inferred one neuron at a time in these data, fluorescent measurements from each neuron are not always of high quality. As a result, activity events cannot be reliably inferred from all neurons \cite{MPPJM15}. Additionally, more neurons could have been observed with less data per neuron if phase tuning was estimated more efficiently. Therefore we applied our robust and scalable Bayesian information sharing algorithm to these data in an attempt to reduce measurement noise, and decrease the required data necessary to attain precision tuning map measurements.

In this setting, let us introduce some simplifying notation. We use $\ell_i$ to denote the total number of spikes that neuron $i$ has fired. As mentioned earlier, $\ell_i \sim 70$ here. Furthermore, we use $\theta_{i,\ell}$ to denote the $\ell$th phase at which neuron $i$ has fired a spike. Then we convert this phase $\theta_{i,\ell}$ to $\bm{y_{i,\ell}}:=[\cos(\theta_{i,\ell}) \sin(\theta_{i,\ell})]'$, a point on the unit circle. 

We model the neuron's tendency to spike at phases that are concentrated around a certain angle using a two dimensional vector $\bm{\beta_i}$. The direction of $\bm{\beta_i}$ is the preferred phase $\theta_i$ and the length of $\bm{\beta_i}$ is the tuning strength $r_i$. If the neuron is highly tuned, that is there no variability among phases at which this neuron fires a spike, then $r_i=1$ and $\bm{\beta_i}$ lies on the unit circle. On the other hand, if the neuron is weakly tuned, that is, there is large variability among phases at which this neuron fires a spike, then $r_i \sim 0$. We relate observation $\bm{y_{i,\ell}}$ to the unknown {$\bm{\beta_i}:= r_i[\cos\theta_i \ \sin\theta_i]'$} as follows 
\begin{eqnarray*}
\bm{y_{i,\ell}} \big | \bm{\beta_i},\sigma,\nu_i &\sim& \N(\bm{\beta_i} ,\nu_i^2 \sigma^2 \bm{I}), \quad \text{for } \ \ell=1,\cdots,\ell_i
\end{eqnarray*} 
where $\bm{\beta_i}$ is related to $\theta_i$ (preferred phase)  and $r_i$ (tuning strength) as follows
\begin{eqnarray*}
\theta_i&:=&\arctan \Bigl[ \frac{\beta_{2,i}}{\beta_{1,i}}\Bigr], \\
r_i &:=& \sqrt{\beta_{2,i}^2+\beta_{1,i}^2}.
\end{eqnarray*}
{
There are two points worth mentioning. First, the Gaussian noise model clearly violates the fact that $\{\bm{y_{i,\ell}}\}$ lie on the unit circle, and should therefore be considered a rather crude approximation. Nevertheless, as demonstrated below, this Gaussian likelihood with our prior in (\ref{eq:beta_prior}), is remarkably effective in estimating the preferred phases $\{ \theta_i \}$ with as little as one observed phase per neuron. Second, the vector representation of the $\ell_i$ spikes that neuron $i$ has fired
$$
\bm{y_i}=
 \begin{bmatrix}
        \bm{y_{i,1}}  \\
       \vdots  \\
        \bm{y_{i,\ell_i}}
     \end{bmatrix}_{2\ell_i \times 1}
$$ can be related to the unknown $\bm{\beta_i}$  using the formulation presented in equation (\ref{eq:model}) where 
$$
\bm{X_i}=
 \begin{bmatrix}
        \vdots  \\
     \bm{I_{2\times 2}} \\
       \vdots  \\
     \end{bmatrix}_{2\ell_i \times 2}.
$$
}

The ML estimate  of $\bm{\beta_i}$, given the Gaussian additive noise model, is the sample mean of the observations $\{\bm{y_{i,\ell}}\}_{\ell=1,\cdots,\ell_i}$,  $\bm{\beta_{i,\text{ml}}} =  \frac{1}{\ell_i} \sum_{\ell=1}^{\ell_i} \bm{y_{i,\ell}} $. The ML estimate of the preferred phase $\theta_{i,\text{ml}} = \arctan \Bigl[ \frac{\beta_{2,i,\text{ml}}}{\beta_{1,i,\text{ml}}}\Bigr]$ is the circular mean of the observed phases, as depicted in figure ~\ref{fig:raw_single}. The resulting radius $\|\bm{ \beta_{i,\text{ml}}} \|_2$, the ML estimate of $r_i$, will be 1 if all angles are equal. If the angles are uniformly distributed on the circle, then the resulting radius will be 0, and there is no circular mean. The radius measures the concentration of the angles and can be used to estimate confidence intervals. 

In addition to the observed phases, we also have the three-dimensional physical location of all cells. As an illustrative example, the spatial distribution of $\{\theta_{i,\text{ml}} \}$ and $\{r_{i,\text{ml}} \}$ is depicted in figure \ref{fig:raw_pp} and \ref{fig:raw_ts}. The three-dimensional location is projected into the two-dimensional x-y plane. Each dot is a cell, and its color in panel \ref{fig:raw_pp} and \ref{fig:raw_ts}  corresponds to  $\theta_{i,\text{ml}}$ and $r_{i,\text{ml}}$, respectively. Clearly, nearby cells tend to have similar preferred phases and tuning strengths --- but there are many exceptions to this trend. A mixture prior is required to avoid oversmoothing the border between clusters of cells with similar properties while allowing cells within a cluster to share information and reduce noise.

\begin{figure}
\vspace{-1cm}
        \centering
        \begin{subfigure}[b]{0.4\textwidth}
                \includegraphics[width=\textwidth]{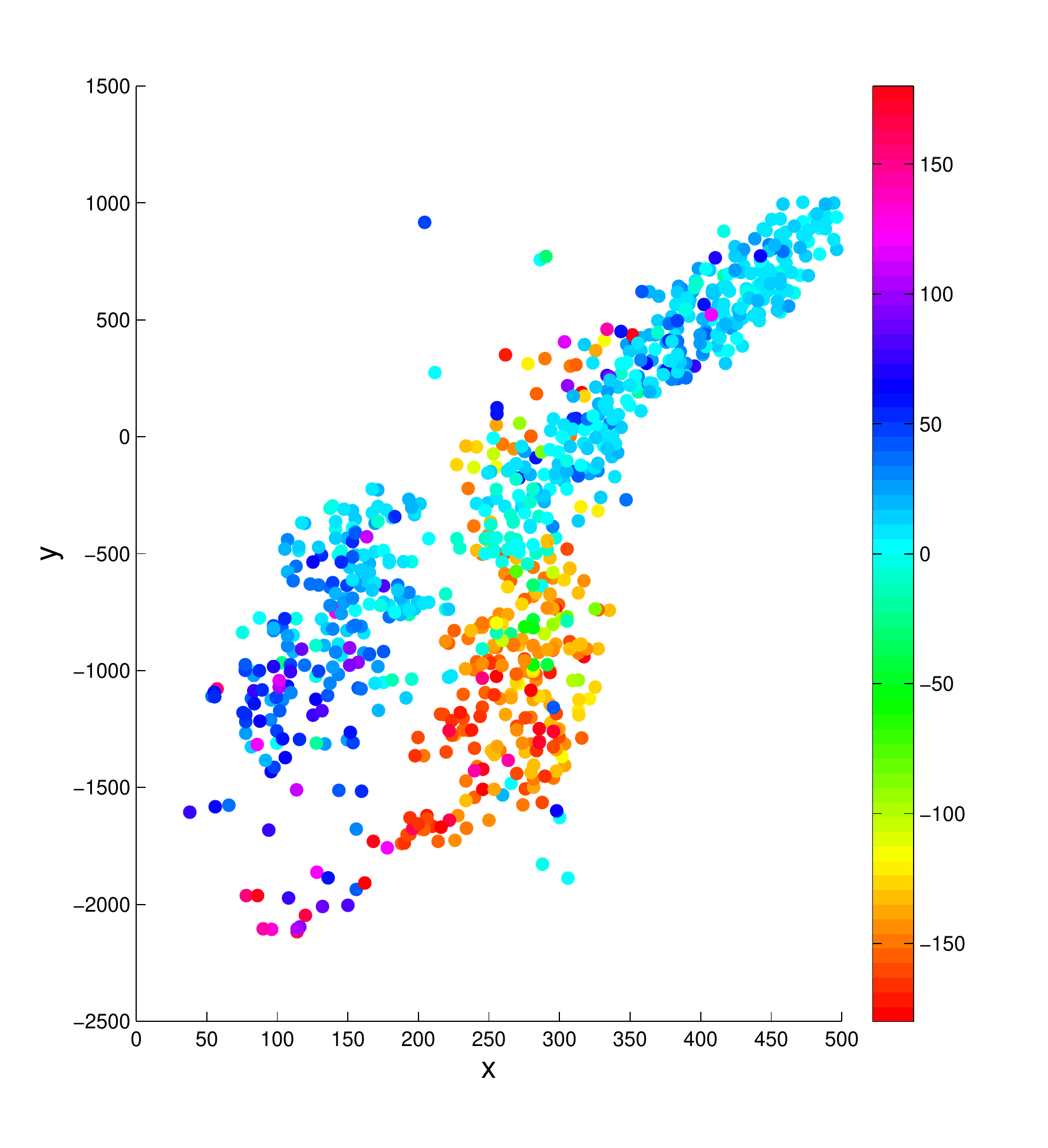}
                \caption{$\lambda=0$}
                \label{fig:raww}
        \end{subfigure}%
        ~ 
        \begin{subfigure}[b]{0.4\textwidth}
                \includegraphics[width=\textwidth]{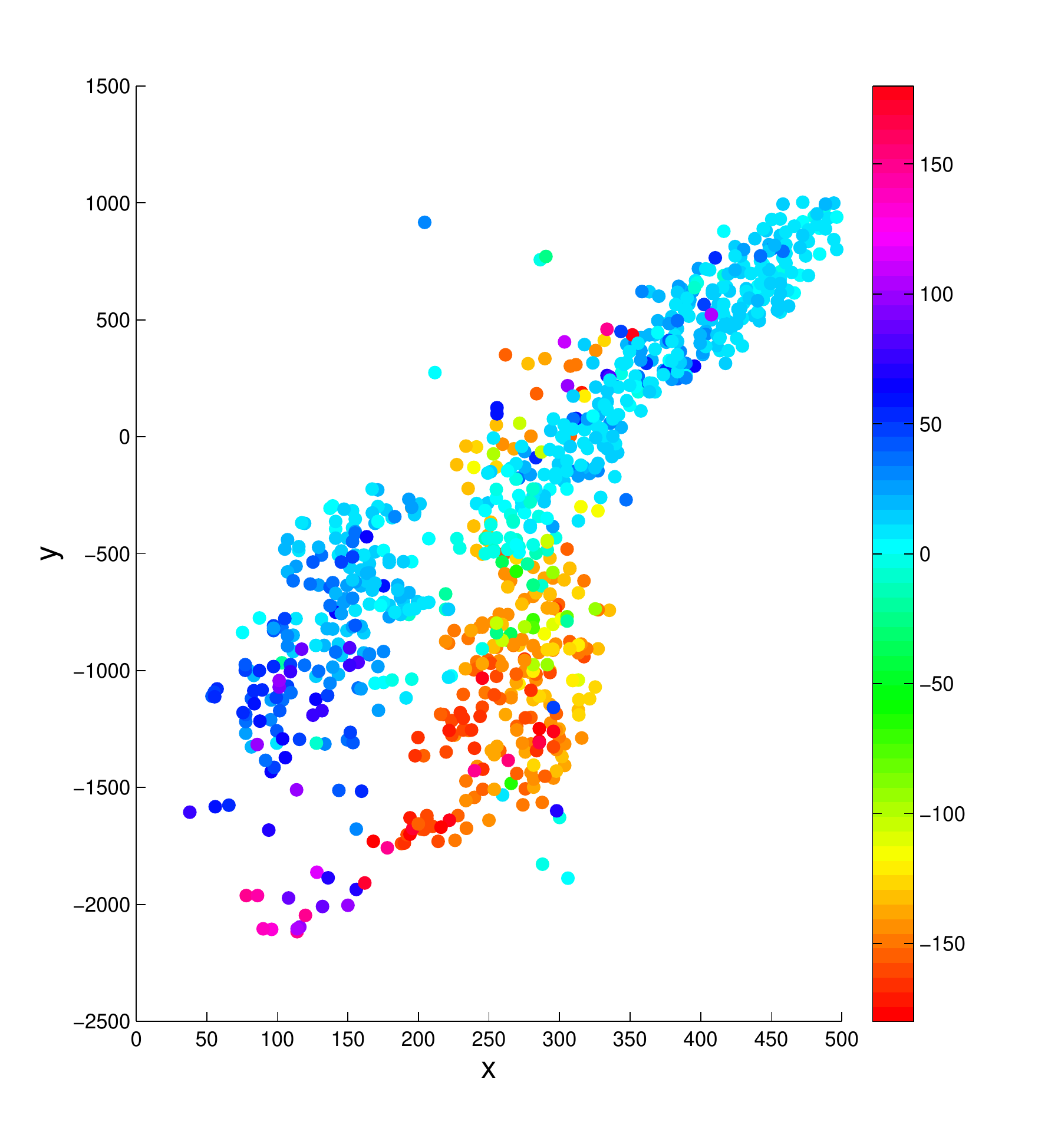}
                \caption{$\lambda=1$}
                \label{fig:lambda_0_0_1}
        \end{subfigure}
          
        \begin{subfigure}[b]{0.4\textwidth}
                \includegraphics[width=\textwidth]{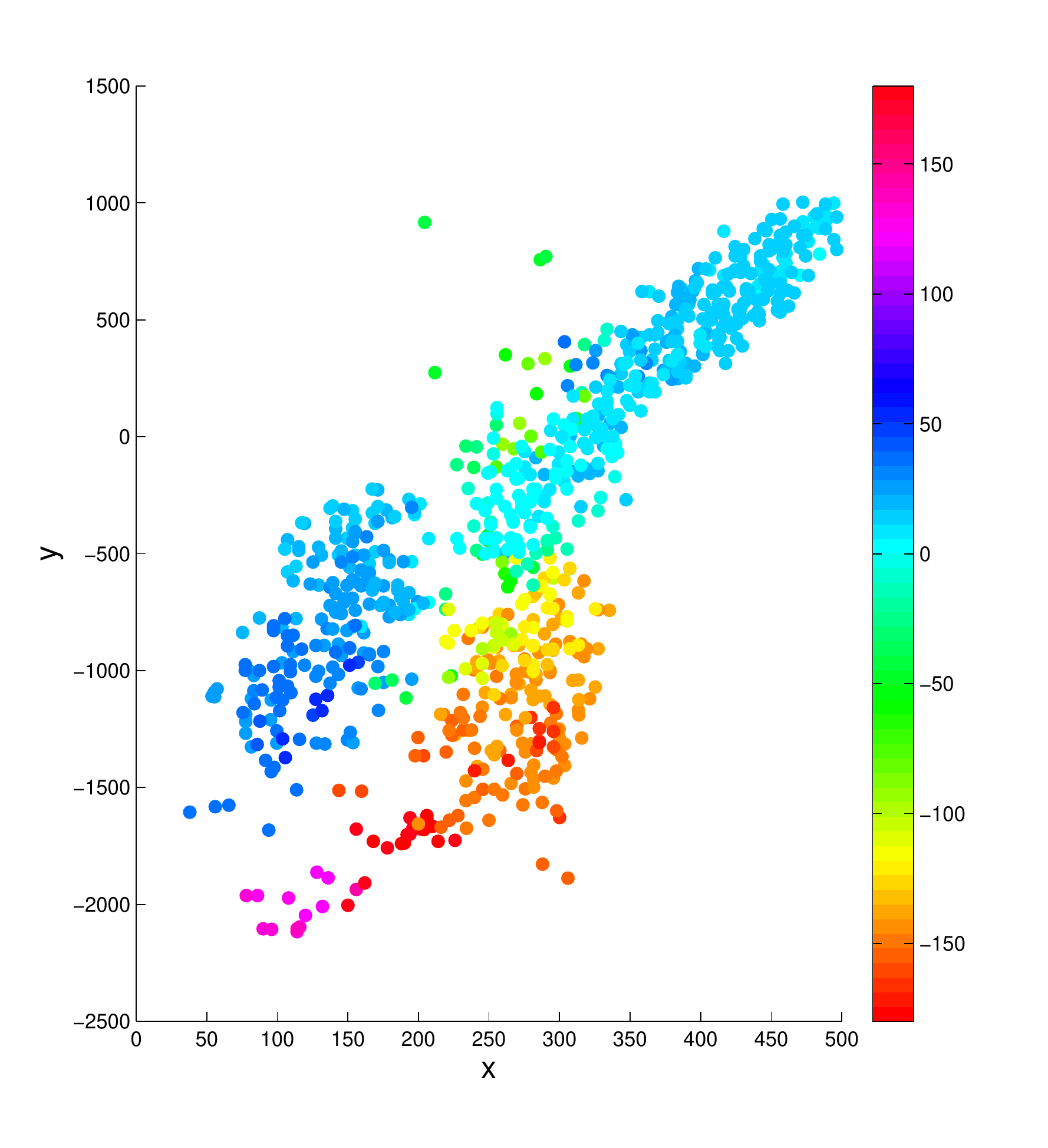}
                \caption{$\lambda = 10$}
                \label{fig:lambda_0_1}
        \end{subfigure}
           ~ 
        \begin{subfigure}[b]{0.4\textwidth}
                \includegraphics[width=\textwidth]{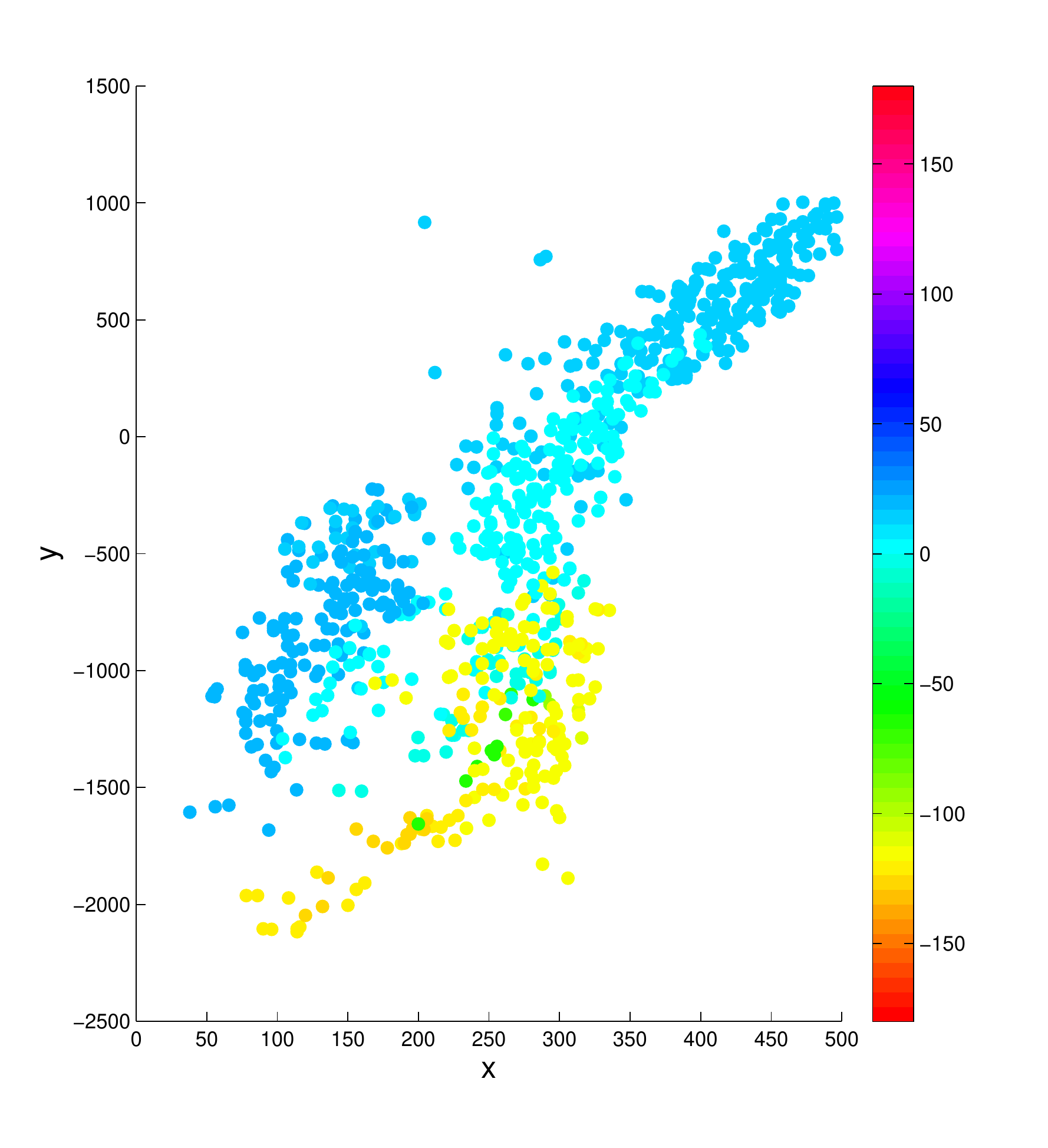}
                \caption{$\lambda=100$}
                \label{fig:lambda_1}
        \end{subfigure}
        \caption{Preferred phase estimates for different values of the hyper parameter $\lambda$.  Each dot corresponds to the estimated preferred angle $\hat \theta_i$ for one cell.  For $\lambda=0$, the estimates are equal to the ML estimates. For $\lambda=1$, information sharing is not large enough and estimates are not very different from the ML estimates. For $\lambda=10$, nearby neurons are forced to have similar preferred phases, nonetheless, the sharp border between functionally different clusters of neurons is not oversmoothed. The posterior mean and standard deviation of $\lambda$, based on 10000 iterations (after 500 burn-ins), is $5.26$ and $0.52$, respectively. For $\lambda=100$, smoothing within clusters is stronger and borders are not
violated. However, tuning estimates within each cluster suffer from oversmoothing.    
           }\label{fig:animals}
\end{figure}

In order to include the physical location of the cells into our Bayesian formulation, we formed a proximity network based on nearest spatially-whitened neighbors, as described in section \ref{sec:ldn}. $\{\bm{\hat \beta_i}\}$, the posterior expectation of $\{\bm{\beta_i}\}$, is based on 10000 samples from our efficient Gibbs sampler (after 500 burn-in iterations). For illustration purposes, we experimented with holding the hyperparamter $\lambda$ fixed in the simulations; the effects of this hyper parameter on the estimates of the preferred phases and tuning strengths
\begin{eqnarray}
\hat \theta_i &:=& \arctan \Bigl ( \frac{   \hat \beta_{2,i}} {\hat \beta_{1,i} } \Bigr),\label{eq:hat_mle}\\
\hat r_{i} &:=& \| \bm{ \hat \beta_i} \|_2\label{eq:hat_mle},
\end{eqnarray}
are depicted in figure \ref{fig:animals}. It is clear that large $\lambda$ forces nearby neurons to have more similar preferred phases whereas for small $\lambda$ the preferred phases revert to their respective ML estimates.

\begin{figure}
        \centering
          \begin{subfigure}[b]{1\textwidth}
                \includegraphics[width=\textwidth]{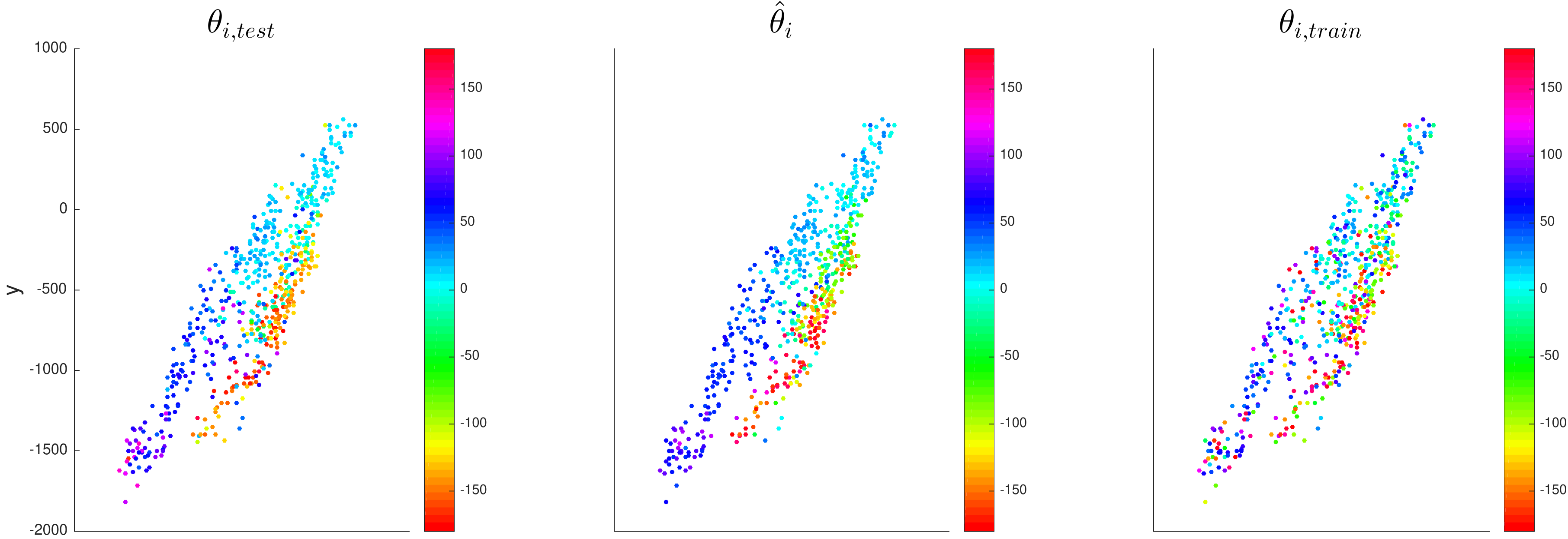}
        \end{subfigure}

          \vspace{1cm}
          

        \begin{subfigure}[b]{1\textwidth}
                \includegraphics[width=\textwidth]{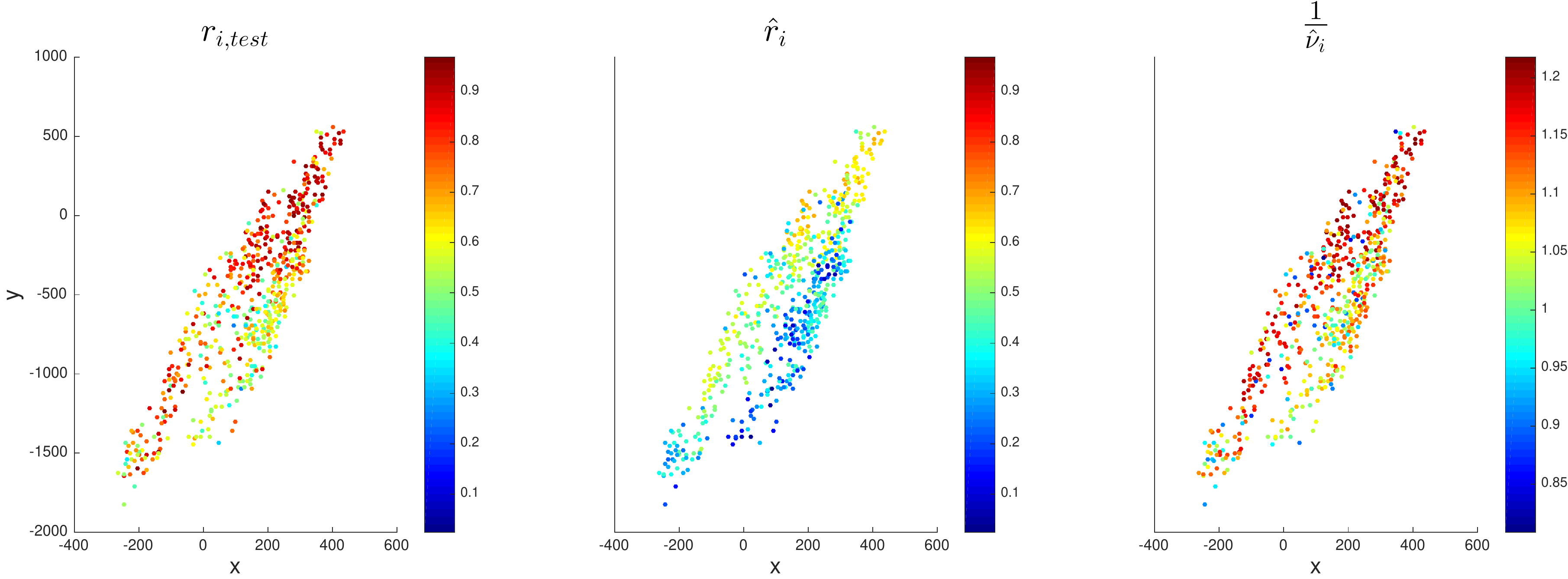}
        \end{subfigure}
        \caption{Dataset 1 with $n=584$. The posterior estimates $\hat \sigma=0.62 \pm 0.02$  and $\hat \lambda = 6.43 \pm 0.38$ (i.e., the mean $\pm$ standard deviation) are based on 10000 samples (after 500 burn-ins). The test set is made of $59$ observed phases per neuron. The test error is $\frac{1}{n} \sum |\hat \theta_i - \theta_{i,\text{test}}|=27.6^\circ$ and the raw error is $\frac{1}{n} \sum |\hat \theta_{i,\text{train}} - \theta_{i,\text{test}}|=36.9^\circ$.  }\label{fig:ttph1}
\end{figure}

\begin{figure}
        \centering
          \begin{subfigure}[b]{1\textwidth}
                \includegraphics[width=\textwidth]{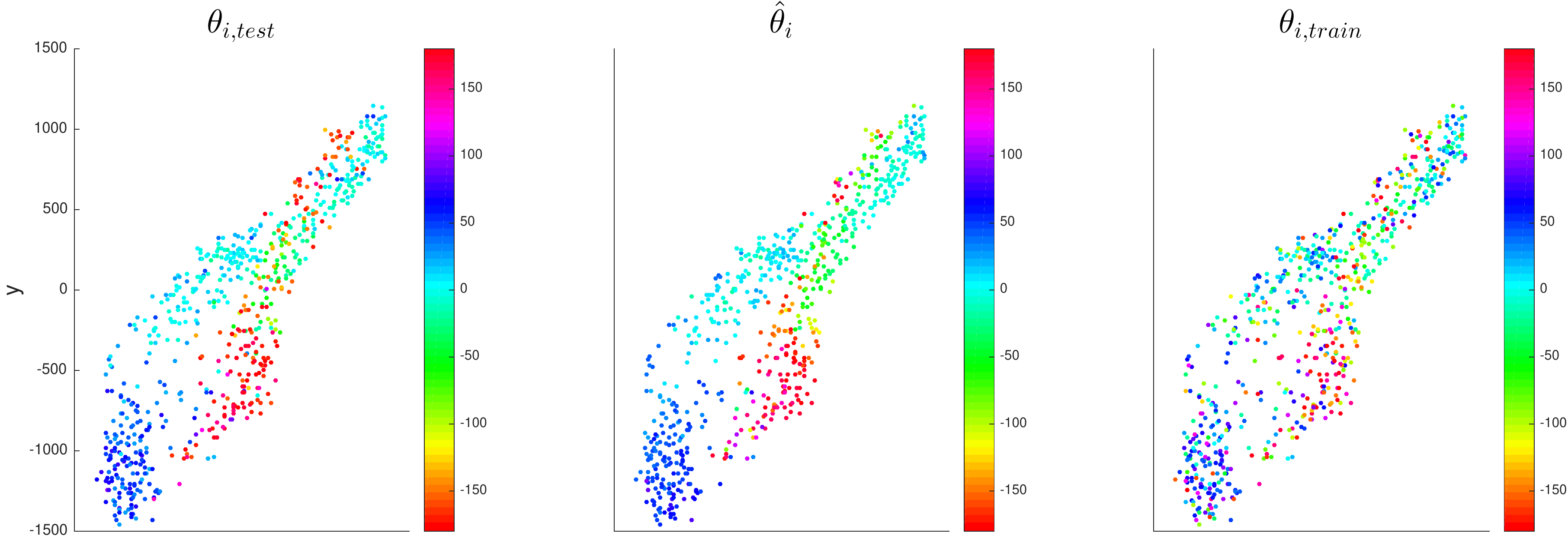}
        \end{subfigure}

          \vspace{1cm}

        \begin{subfigure}[b]{1\textwidth}
                \includegraphics[width=\textwidth]{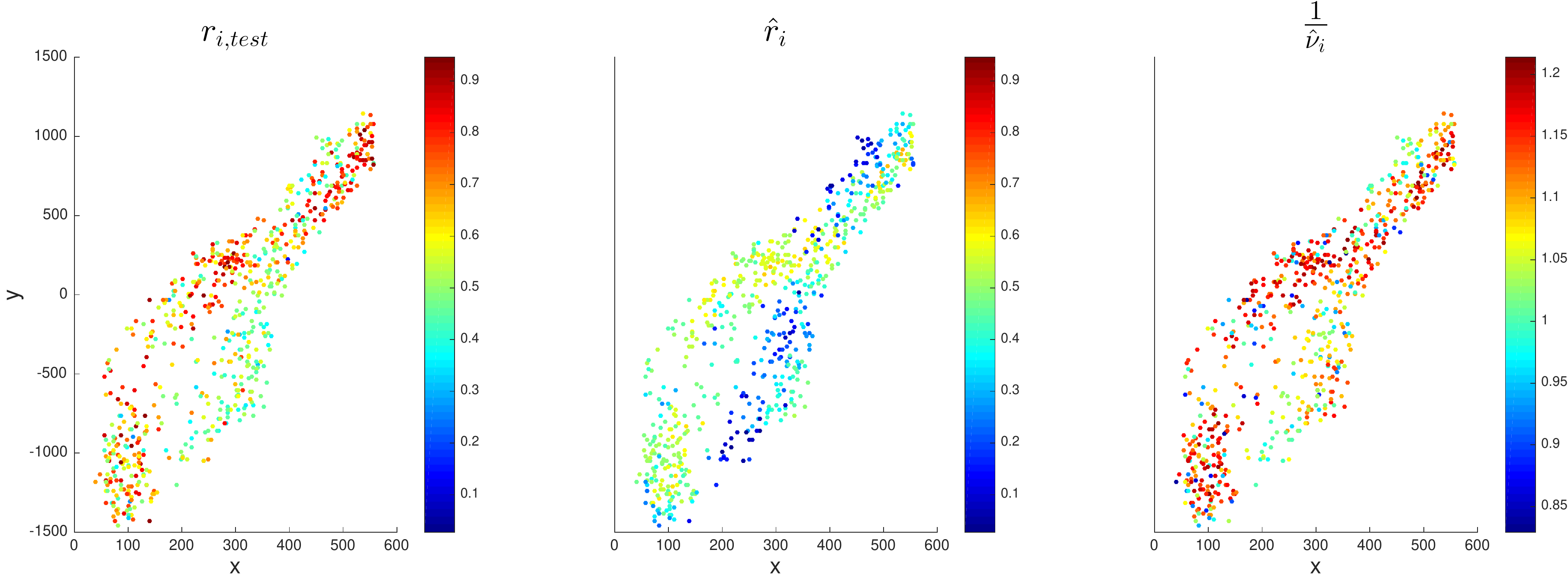}
        \end{subfigure}
        \caption{Dataset 16 with $n=676$. The posterior estimates $\hat \sigma=0.62 \pm 0.01 $  and $\hat \lambda = 7.04 \pm 0.38$ (i.e., the mean $\pm$ standard deviation) are based on 10000 samples (after 500 burn-ins). The test set is made of $60$ phases per neuron. The test error is $\frac{1}{n} \sum |\hat \theta_i - \theta_{i,\text{test}}|=28.4^\circ$ and the raw error is $\frac{1}{n} \sum |\hat \theta_{i,\text{train}} - \theta_{i,\text{test}}|=41.5^\circ$.  }\label{fig:ttph16}
\end{figure}

\begin{figure}
        \centering
          \begin{subfigure}[b]{1\textwidth}
                \includegraphics[width=\textwidth]{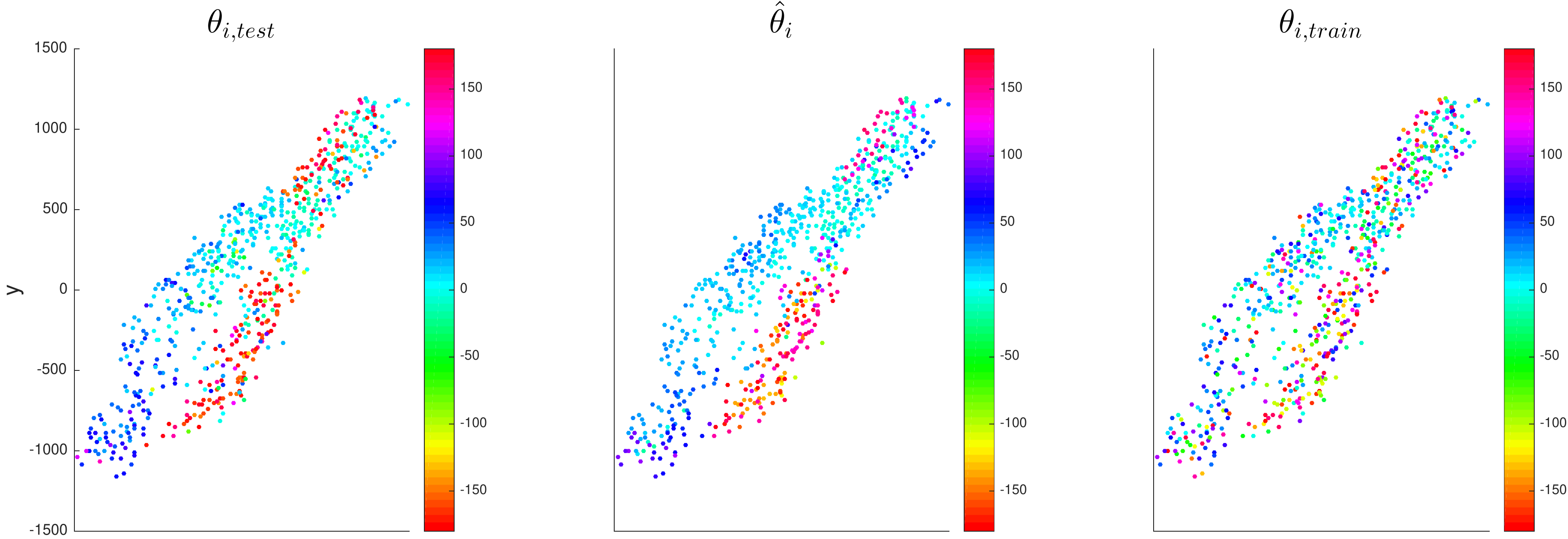}
        \end{subfigure}

          \vspace{1cm}

        \begin{subfigure}[b]{1\textwidth}
                \includegraphics[width=\textwidth]{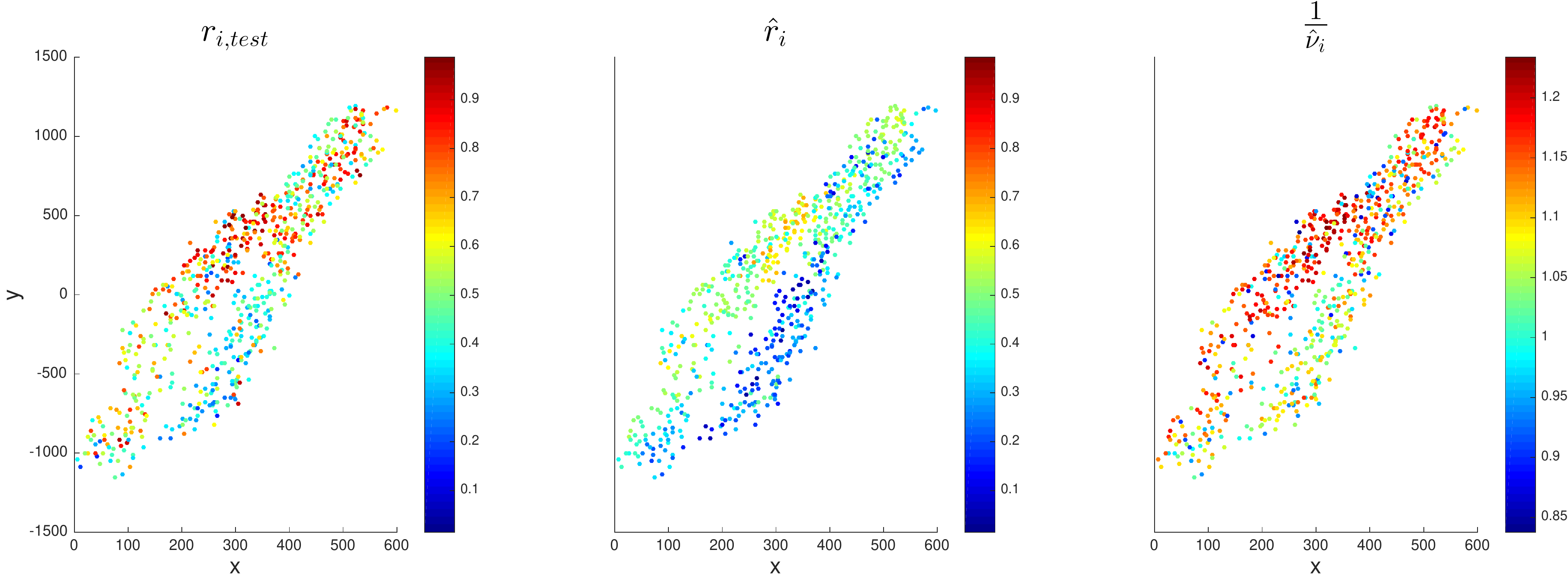}
        \end{subfigure}
        \caption{Dataset 23 with $n=695$. The posterior estimates  $\hat \sigma=0.69 \pm  0.02$  and $\hat \lambda = 7.39 \pm 0.39$ (i.e., the mean $\pm$ standard deviation) are based on 10000 samples (after 500 burn-ins). The test set is made of $73$ phases per neuron. The test error is $\frac{1}{n} \sum |\hat \theta_i - \theta_{i,\text{test}}|=39.45^\circ$ and the raw error is $\frac{1}{n} \sum |\hat \theta_{i,\text{train}} - \theta_{i,\text{test}}|=51.55^\circ$.  }\label{fig:ttph23}
\end{figure}

\begin{figure}
        \centering
          \begin{subfigure}[b]{1\textwidth}
                \includegraphics[width=\textwidth]{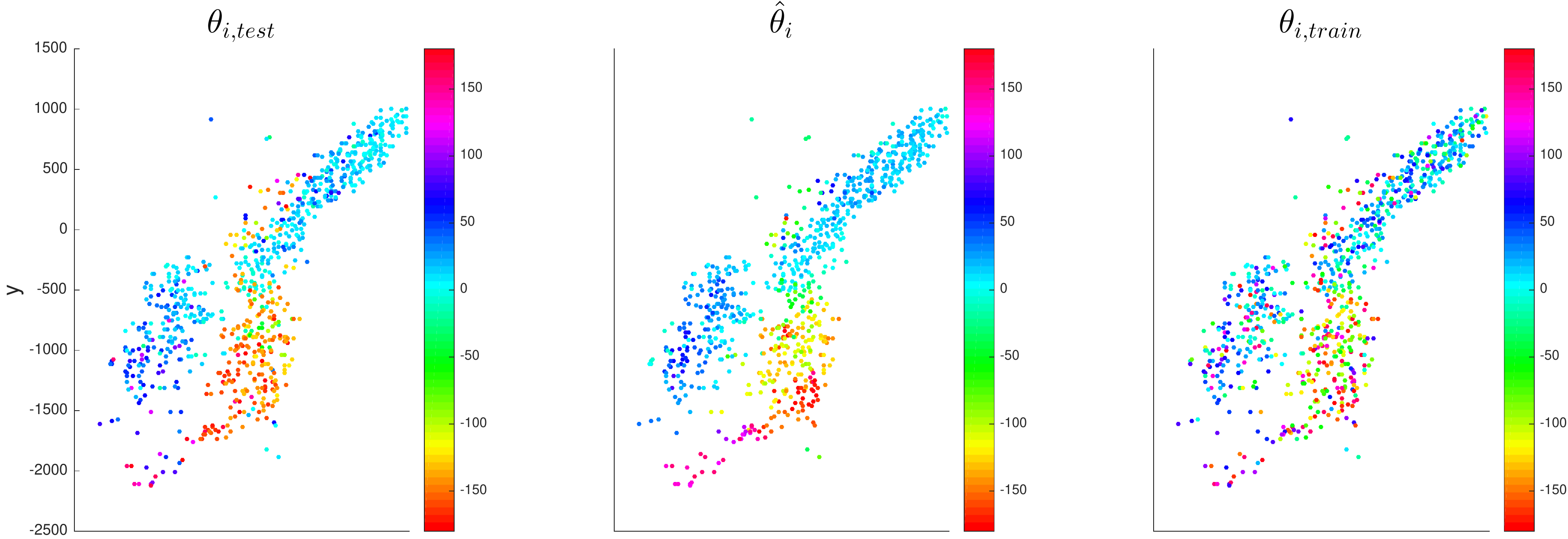}
        \end{subfigure}

          \vspace{1cm}

        \begin{subfigure}[b]{1\textwidth}
                \includegraphics[width=\textwidth]{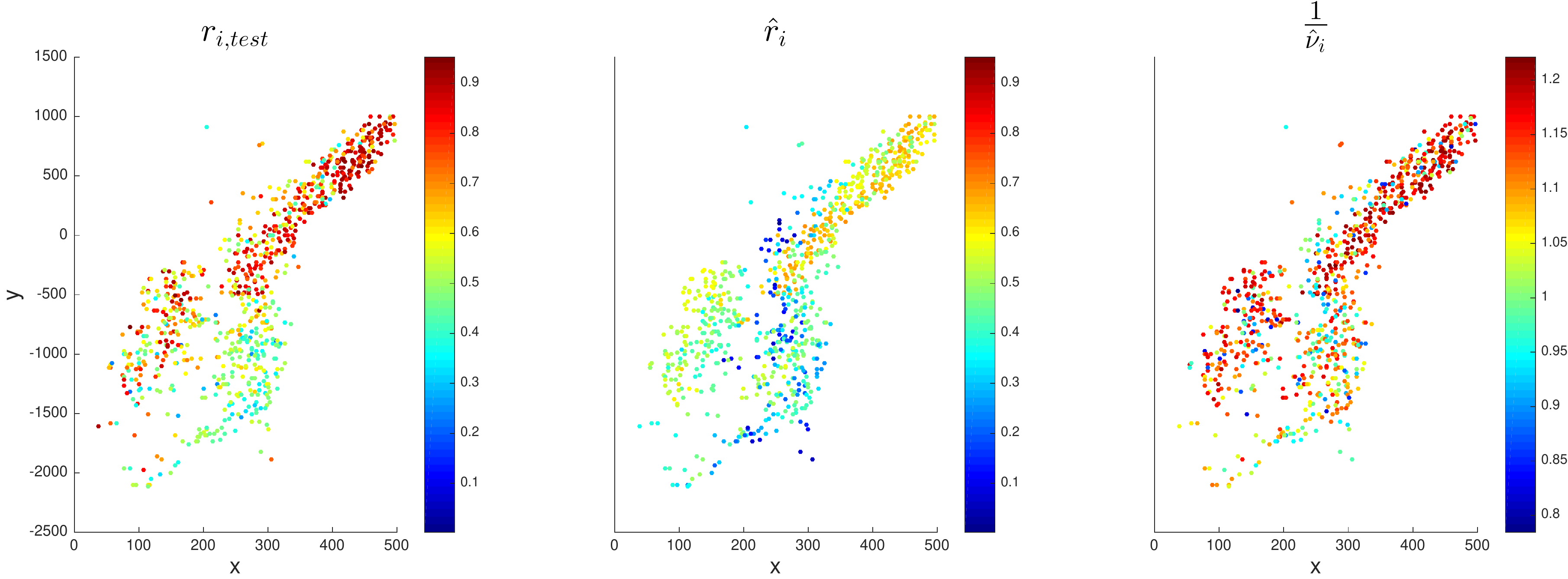}
        \end{subfigure}
        \caption{Dataset 26 with $n=854$. The posterior estimates  $\hat \sigma= 0.62 \pm 0.01$  and $\hat \lambda =7.52 \pm 0.40$ (i.e., the mean $\pm$ standard deviation) are based on 10000 samples (after 500 burn-ins). The test set is made of $82$ phases per neuron. The test error is $\frac{1}{n} \sum |\hat \theta_i - \theta_{i,\text{test}}|=27.1^\circ$ and the raw error is $\frac{1}{n} \sum |\hat \theta_{i,\text{train}} - \theta_{i,\text{test}}|=42.9^\circ$.  }\label{fig:ttph26}
\end{figure}

The ability of our method to recover the preferred phases from as little as one noisy phase $\theta_{i,\ell}$ per neuron is illustrated below. We divide the data into two parts. For each cell, there are roughly 70 phases recorded (at which the corresponding neuron fired). For each neuron $i$, we randomly selected one of the phases $\{\theta_{i,\ell} \}_{\ell=1,\cdots,\ell_i}$ for the training set, and let the rest of the phases constitute the testing set:
\begin{eqnarray*}
\bm{y_{i,\text{train}} }&:=&  \begin{pmatrix}
\cos(\theta_{i,\ell_{\text{train}}})\\
\sin(\theta_{i,\ell_{\text{train}}})
\end{pmatrix} ,
\quad
\bm{y_{i,\text{test}}} := \frac{1}{\ell_i-1} \sum_{\substack{\ell=1,\cdots,\ell_i \\  \ell \neq \ell_{\text{train}} }}  \begin{pmatrix}
\cos(\theta_{i,\ell})\\
\sin(\theta_{i,\ell})
\end{pmatrix}. \label{eq:testphase}
\end{eqnarray*}
The raw estimates of preferred phases and tuning strengths, using training data, are computed as follows{
\begin{eqnarray*}
\theta_{i,\text{train}} &:=& \arctan \Bigl ( \frac{   y_{2,i,\text{train}}} {y_{1,i,\text{train}} } \Bigr),
\quad r_{i,\text{train}} := \|   \bm{y_{i,\text{train}}} \|_2, 
\end{eqnarray*}}
 and raw estimates of preferred phases and tuning strengths, using testing data, are computed likewise. For $\{\bm{y_i}\}$ in our Gibbs sampler, we use the training data $\{\bm{y_{i,\text{train}}} \}$. Posterior estimates $\{\hat \theta_i,\hat r_i, \hat \nu_i \}$, for four distinct datasets, are depicted and compared against testing data in figures  
 \ref{fig:ttph1}-\ref{fig:ttph26}.  For $\lambda^2$, we use a Gamma($r=1, \delta=1$) prior and for $\sigma^2$ an improper inverse-Gamma($\kappa'=0$, $\epsilon'=0$) prior. Generally speaking, $\sigma$ and $\lambda$ are not identifiable. Furthermore, the joint posterior distribution of $\beta$ and $\sigma$ is only unimodal given $\{\nu_i^2 \}$. We address both challenges by placing a relatively tight prior on $\{\nu_i^2 \}$. We use independent $\text{inverse-Gamma}(\varkappa=3, \varepsilon=2)$ priors for $\{\nu_i^2 \}$, making the prior means and variances equal to one. Since the posterior distribution of $\beta$ and $\sigma$ is only unimodal given $\{\nu_i\}$, this prior constrains the $\nu_i$s such that the posterior distribution stays nearly unimodal. Finally, $\lambda$ and $\sigma$ stay nearly identifiable given this tight prior. 
 
 
The raw training estimates of preferred phases and tuning strengths are very noisy which is expected given the fact that only one phase per neuron is used. This is an extremely low signal-to-noise limit. In contrast, roughly 70 phases per neuron are used to compute the raw testing estimates. The Bayesian estimates $\{\hat \theta_i,\hat r_i \}$ are also based on one phase per neuron, but they employ the a priori knowledge that the activity of a neuron carries information about its nearby neurons. As mentioned earlier, this is done by incorporating the proximity network into the Bayesian formulation.

Moreover, as illustrated in the middle panels of figures \ref{fig:ttph1}-\ref{fig:ttph26}, the Bayesian estimates respect the border of clustered cells with similar phasic preferences and tuning strengths. Information is not invariably shared among nearby cells; instead, it is based on how locally similar the samples of $\{\bm{\beta_i}\}$ are. If  the estimated typical noise is much less than the local difference, then, intuitively speaking, local smoothing should be avoided because the difference seems statistically significant.

\begin{figure}
    \includegraphics[width=1\textwidth]{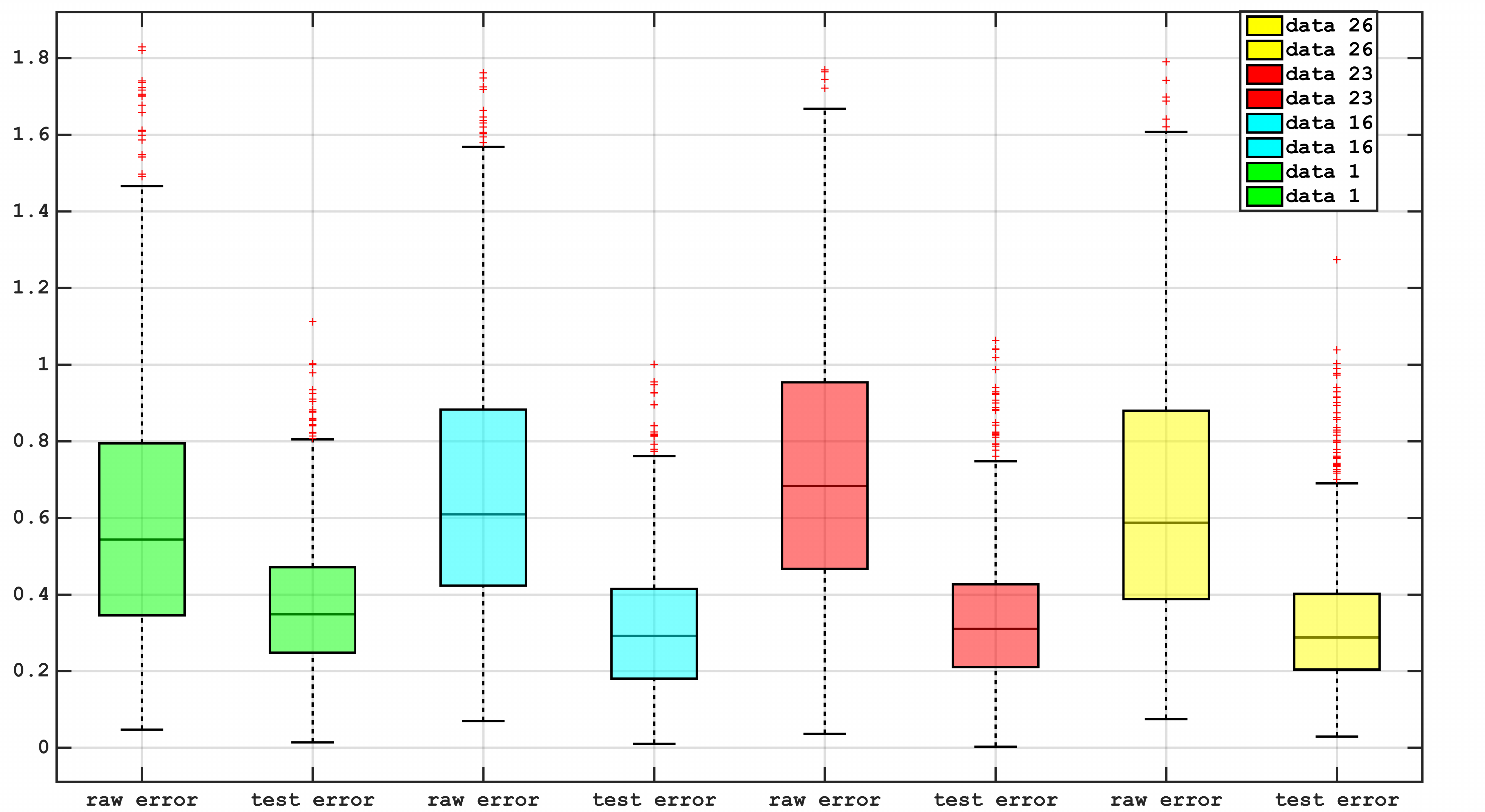}
  \caption{ Tukey boxplots comparing the raw error and test error for the four datasets  illustrated in figures (\ref{fig:ttph1}-\ref{fig:ttph26}). The raw and test error (for cell $i$) are defined as $\|\bm{ y_{i,\text{train}} -y_{i,\text{test}}} \|_2$ and $\| \bm{\hat \beta_{i} -y_{i,\text{test}} }\|_2$, respectively. 
  } \label{fig:summary}
\end{figure}

In contrast, the raw test estimates $\{\theta_{i,\text{test}}, r_{i,\text{test}} \}$ are computed in isolation (one neuron at a time) but use roughly 70 phases per neuron (high signal-to-noise). The Bayesian estimates are less noisy in comparison to the raw training estimates (low signal-to-noise) and qualitatively resemble the raw test estimates (high signal-to-noise). Unlike the previous  synthetic data example, here the true parameters are unknown. In order to quantify the noise reduction, we treat the high signal-to-noise test estimates as the unknown true parameters, and compare them against the Bayesian estimates. Recall that the Bayesian estimates are  based on the low signal-to-noise raw train data. We quantify the noise reduction, by comparing the testing error $\frac{1}{n} \sum |\hat \theta_i - \theta_{i,\text{test}}|$ with the raw error $=\frac{1}{n} \sum | \theta_{i,\text{train}} - \theta_{i,\text{test}}|$. The test error is $10^\circ-16^\circ$ less than the raw error. For more details see the the captions of figures \ref{fig:ttph1}-\ref{fig:ttph26}. Lastly, the boxplots in figure \ref{fig:summary} summarize and quantify the noise reduction due to our robust Bayesian information sharing approach.  In each case, the new Bayesian approach provides significant improvements on the estimation accuracy.

%


\clearpage
\section{Concluding Remarks\label{sec:cr}}

We developed a robust and scalable Bayesian smoothing approach for inferring tuning functions from large scale high resolution spatial neural activity, and illustrated its application  in a variety of neural coding settings. A large body of work has addressed the problem of estimating a smooth spatial process from noisy observations \cite{B74,Wahba90,BK95,RH05,GPBook06}. These ideas have found many of their applications in problems involving tuning function estimation \cite{GAO02,Cunningham07,CZANNER05,CGRS09, PAFKRVVW10,Kamiar08,MGWKB10,MGWKB11,P10,PRHP14}. { There has also been some work on parametric Bayesian tuning function estimation (see \cite{CSSK10} and references therein).} The main challenge in the present work was the large scale (due to the high spatial resolution) of the data and the functional discontinuities present in neuronal tuning maps, e.g.  \cite{SG96,OHKI05,MPPJM15}.


In order to address these challenges, we proposed a robust prior as part of a computationally efficient  block Gibbs sampler that employs fast Gaussian sampling techniques \cite{HR91,H09,PaYu10} and the Bayesian formulation of the Lasso problem \cite{PG08,CGGK10}.  This work focused especially on the conceptual simplicity and computational efficiency of the block Gibbs sampler: we emphasized the robustness properties of the Bayesian Lasso, the unimodality of the posterior and the use of efficient linear algebra methods for sampling, which avoid the Cholesky decomposition or other expensive matrix decompositions. Using in vitro recordings from the spinal cord, we illustrated that  this approach can effectively infer tuning functions from noisy observations, given a negligible portion of the data and reasonable computational time.

\comment{

In a related line of work, relevant vector machines allow the degree of smoothness to be uneven and infer it from local statistics \cite{TIP01,Bishop06}. This is typically achieved using the following prior,
\begin{eqnarray*}
p(\beta| \{ w_{ij} \}) &\propto&  \prod_{i \sim j} \exp \bigl ( - \frac{w_{ij}}{2 } \| \beta_i - \beta_j\|_2^2 \bigr)
\end{eqnarray*}
The values of the local smoothing parameters $\{w_{ij} \}$ can be determined using evidence approximation, in which the marginal likelihood is maximized. Since optimization of the marginal likelihood is intractable, coordinate descent algorithms have been proposed \cite{FT01}. The computational cost per spatial point scales quadratically with total number of spatial points, amounting to a computational cost  that scales cubically with the total number of spatial points, making it computationally infeasible for high dimensional datasets. Finally, local maxima in the model's marginal likelihood function surface can be a significant concern in some cases. 
}

It is worth mentioning that in another line of work, smoothness inducing priors were used to fit spatio-temporal models to fMRI data \cite{PTF05, GCW09,HG10,QDG10, W12}. Although these priors handle spatial correlation in the data, they do not always successfully account for spatial discontinuities and the large scale of the data. \cite{WJBS04} used automatic relevance determination (ARD) \cite{M95} to allow for spatially non-stationary noise where the level of smoothness at each voxel was estimated from the data. It is known \cite{WN08} that ARD can converge slowly to suboptimal local minima. On the other hand, wavelet bases with a sparse prior, defined by a mixture of two Gaussian components, allowed \cite{GP07}  to present a statistical framework for modeling  transient, non-stationary or spatial varying phenomenon. They used variational Bayes approximations together with fast orthogonal wavelet transforms to efficiently compute the posterior distributions. As mentioned in their paper, a main drawback is that  wavelet denoising with an orthogonal transform exhibits Gibbs phenomena around discontinuities, leading to inefficient modeling of singularities, such as edges. In \cite{VCDH10,SZT10, S12,GKKKT13, HWRGBJS15}   smoothness (and matrix factorization) approaches  were combined with various global sparsity-inducing priors (or regularizers) to smooth (or factorize) the spatio-temporal activity of voxels  that present significant effects, and to shrink to zero voxels with insignificant effects. In \cite{HPATF07}, non-stationary Gaussian Processes were used as adaptive filters with the computational disadvantage of inverting large covariance matrices. Finally, in a recent work, \cite{SEBV16} design an efficient Monte Carlo sampler to perform spatial whole-brain Bayesian smoothing. Costly Cholesky decompositions are avoided by efficiently employing the sparsity of precision matrices and  preconditioned conjugate gradient methods. The prior in  \cite{SEBV16} assigns a spatially homogenous level of smoothness which performs less favorably in situations involving outliers and sharp breaks in the functional map.

{
There is also a vast literature addressing the recovery of images from noisy observations (see \cite{MGMH04,BCM05} and references therein). Most of these techniques use some sort of regularizer or prior to successfully retain image discontinuities and remove noise.

Early examples include the auxiliary line process  based quadratic penalty in \cite{GG84}, and the $\sum_i \frac{1}{1+|\nabla_i \bm{\beta}|}$ log-prior in \cite{GR92}, where  the gradient at  $i$ is denoted by $\nabla_i$. The line process indicates sharp edges and suspends or activates the smoothness penalty associated with each edge.  The log-prior  $\sum_i \frac{1}{1+|\nabla_i \bm{\beta}|}$ encourages the recovery of discontinuities while rendering  auxiliary variables of the line process as unnecessary. These log-priors are non-concave. These non-concave maximum a posteriori optimization problems are generally impractical to maximize. Different techniques were designed based on  simulated annealing \cite{GG84,GR92,GY95}, coarse-to-fine optimization \cite{BL88} and alternate maximization between image and  auxiliary contour variables \cite{CBAB95}  to compute (nearly) global optimums, at the expense of prohibitively large amount of computation. Moreover, it is known that a small perturbation in the data leads to abrupt changes in the de-noised image \cite{BS93}. This is due to the non-concavity of the problem.

Image de-noising methods based on concave log-priors (or regularizers)  that enjoy edge preserving properties were designed in  \cite{SD90,G90,SD91,ROF92,BS93}. These log-priors typically take the form of $-\sum_i \phi(\nabla_i \bm{\beta})$ for some concave function $\phi(.)$. Example of $\phi(x)$ include the Huber function \cite{H64}  in \cite{SD90,SD91}, $\log \cosh (\frac{ x }{T})$  in \cite{G90},  $ |x |^p$ where $1 \leq p < 2$  \cite{BS93} and  $ |x|$  in the  TV penalty \cite{ROF92,B93}. Various methods have been proposed for computing optimal or nearly optimal solutions to  these image recovery problem, e.g. \cite{SD90,G90,SD91,ROF92,BS93,VO95,VO98,C04,WYYZ08,OBF09,ABF10,BS11,DVL11}.

In addition to these approaches, another significant contribution has been to consider  wavelet, ridgelet and curvelet  based priors/regularizers, e.g. \cite{DJ94,C99,EC99,SCD02,PSWS03}  which present  noticeable improvements in image reconstruction problems. More recently, the non-local means method \cite{BCM05,DFKE07,LBM13} presents a further improvement. However, most of these methods' favorable performance relies heavily on parameters which have been fine tuned for specifically additive noisy observations of two-dimensional arrays of pixels of real world images. In other words, they are specifically tailored for images. It is not clear if and how these approaches can be modified to retain their efficiency while being applied to broader class of spacial observations lying on generic graphs. Moreover, the denoised images rarely come equipped with confidence intervals.   But our sampling based approach allows for proper quantification of uncertainty, which could in turn be used to guide online experimental design; for example, in the spinal cord example analyzed here, we could choose to record more data from neurons with the largest posterior uncertainty about their tuning functions.
 
In principle, many of above mentioned approaches can be formulated as Bayesian, with the aid of the Metropolis-Hastings (MH) algorithm,  to compute posterior means and standard deviations \cite{LS04,LM13}. However, generic MH approaches can lead to unnecessary high computational cost. For example, in \cite{LS04} a TV prior and Gaussian noise model was used to denoise a one dimensional pulse; it was reported that  the chain resulting from the MH algorithm suffers from very slow convergence. One contribution of the present paper is to show that by using a hierarchical representation of our prior in equation (\ref{eq:hie})   costly MH iterations can be avoided in all steps of our block Gibbs sampler. Additionally, we show how our model can take into account nonuniform noise variance (quite common in neuroscience applications) without increasing the computational complexity. Finally, we emphasize the importance of conditioning  $\bm{\beta}$ on  $\sigma$  in equation (\ref{eq:beta_prior}) which has been neglected in previous Bayesian formulations of the TV prior \cite{LS04,LM13}. This is important because it guarantees a unimodal posterior of  $\bm{\beta}$ and $\sigma$ given $\{ \nu_i \}_{i=1,\cdots,n}$ and $\lambda$. 

}

We should also note that a number of fully-Bayesian methods have been developed that present adaptive smoothing approaches for modeling non-stationary spatial data. These methods are predicated on the idea that to enhance spatial adaptivity, local smoothness parameters should  a priori be viewed  as a sample from a common ensemble. Conditional on these local smoothing parameters, the prior is a Gaussian Markov random field (GMRF) with a rank deficient precision matrix \cite{LFF02,LB04,FKL04,RH05,YS10,YLL10,YSS12}. The hyper prior for the local smoothing parameters can be specified in two ways. The simpler formulation  assumes the local smoothing parameters to be independent  \cite{LFF02,LB04,FKL04,BFH07}. For example, \cite{LFF02} presented a nonparametric prior for fitting unsmooth and highly oscillating functions, based on a hierarchical extension of  state space models where the noise variance of the unobserved states is locally adaptive. The main computational burden lies on the Cholesky decomposition \cite{BFH07}  or other expensive  matrix decompositions of the precision matrix. In a more complex formulation, the log-smoothing parameters  follow another GMRF on the graph defined by edges $i \sim j$ \cite{YS10,YLL10,YSS12}. 
In both formulations,  local smoothing parameters are conditionally dependent, rendering Metropolis-within-Gibbs sampling necessary. { These methods often provide superior estimation accuracy for functions with high spatial variability on regular one-dimensional and two-dimensional lattices, but at a prohibitively higher computational cost which makes them less attractive for the high dimensional datasets considered in this paper.} One interesting direction for future work would be to combine the favorable properties of these approaches with those enjoyed by our scalable and robust Bayesian method.

{

Finally, important directions for future work involve extensions that allow the treatment of point processes, or other non-Gaussian data, and correlated neural activities. Since our prior can be formulated in a hierarchical manner, when dealing with non-Gaussian likelihoods, it is only step 2 of our Gibbs sampler that needs modification. In step 2, all MCMC algorithms suited for Gaussian priors and non-Gaussian likelihoods can be integrated into  our efficient Gibbs sampler. For example, the elliptical slice sampler \cite{MAM10} or Hamiltonian
Monte Carlo methods \cite{DKPR87,RS03,RC05,APP11,GCC11} are well-suited for sampling from posteriors arising from a Gaussian prior and likelihoods from the exponential family. With regard to correlated neural activities,  it would be interesting to see how tools developed in \cite{VASPKLCSP12,BMS12} can be incorporated into our Gibbs sampler to make inference about models which can account for correlated observations.

}

{
\appendix
\section{Unimodality of the Posterior}\label{sec:app}
Here we demonstrate that the joint posterior of  $\bm{\beta}$ and $\sigma^2$ given $\{ \nu_i \}_{i=1,\cdots,n}$ and $\lambda$  is unimodal under the prior in equation (\ref{eq:beta_prior}) and equation (\ref{eq:hpriors}). Note that our discussion here is very similar to that of \cite{PG08}. The joint prior is
\begin{eqnarray*}
p(\bm{\beta},\sigma^2 | \lambda)  &=& \frac{\epsilon^{\kappa} }{\Gamma(\kappa)} (\sigma^2)^{-\kappa-1} e^{-\epsilon/\sigma^2} \prod_{i \sim j}\bigl ( \frac{\lambda}{2 \sigma} \bigr)^m \exp \Bigl( -\frac{\lambda}{\sigma} \Bigl \|  \bm{\beta_i - \beta_j} \Bigr\|_2 \Bigr). 
\end{eqnarray*}
The log posterior is
\begin{equation}
-\Bigl(\kappa+1 + \frac{nd+pm}{2} \Bigr) \log \sigma^2 -\frac{\epsilon}{\sigma^2} -\frac{\lambda}{\sqrt{\sigma^2}} \sum_{i \sim j} \Bigl \|  \bm{\beta_i - \beta_j} \Bigr\|_2 -\frac{1}{2 \sigma^2}\sum_{i=1}^n \frac{\Bigl \| \bm{  y_i - X_i \beta_i }\Bigr\|_2^2}{\nu_i^2} \label{eq:postt}
\end{equation}
ignoring all the terms independent of $\bm{\beta}$ and $\sigma^2$. The mapping (and its inverse)
\begin{equation}
\bm{\phi_i} \leftrightarrow \frac{\bm{\beta_i}}{\sqrt{\sigma^2}}, \hspace{3cm} \rho \leftrightarrow \frac{1}{\sqrt{\sigma^2}}
\end{equation}
is continuous. Therefore, unimodality in the mapped coordinates is equivalent to unimodality in the original coordinates. The log posterior (\ref{eq:postt}) in new coordinates is
\begin{equation}
\Bigl(2\kappa+2 + nd+pm \Bigr) \log \rho -\epsilon \rho^2 -\lambda \sum_{i \sim j} \Bigl \|  \bm{\phi_i - \phi_j} \Bigr\|_2  -\frac{1}{2}\sum_{i=1}^n \Bigl \| \rho\bm{\underline{y}_i} - \bm{\underline{X}_i \phi_i }\Bigr\|_2^2, \label{postcave}
\end{equation} 
where we have earlier defined in equation (\ref{eq:u})
\begin{eqnarray*}
\bm{\underline y_i} &:=& \frac{\bm{y_i}}{\nu_i}, \quad \bm{\underline X_i} := \frac{\bm{X_i}}{\nu_i}. 
\end{eqnarray*}
The log posterior in equation (\ref{postcave}) is clearly concave in $(\bm{\phi_1},\cdots,\bm{\phi_n},\rho)$, and hence the  posterior is unimodal.
}

\section*{Acknowledgements}
We are grateful to A. Brezger, B. Babadi, J. Jahani, A. Maleki, K. Miller, C. Smith, A. Pakman and R. Yue for fruitful conversations and to M. Schnabel for sharing the code that produced the synthetic orientation maps in figure \ref{fig:orientation_preference}. We are also grateful to the laboratory of Thomas M. Jessell, in the Department of Biochemistry and Molecular Biophysics at Columbia University, for providing the spinal cord data presented in section \ref{sec:motor}. We  also would like to thank the referees for carefully reading the original manuscript and raising important issues which improved the paper significantly.
\bibliographystyle{abbrv}
\bibliography{mybib}
\end{document}